\documentstyle[prb,multicol,aps,epsf]{revtex}
\begin{document}
\title{Magnetically Stabilized Nematic Order I:
\\ Three-Dimensional Bipartite Optical
Lattices}
\author{F. Zhou$^{\dagger}$, M. Snoek$^{\dagger\dagger}$, J. Wiemer$^{\dagger\dagger}$ and I. 
Affleck$^{\dagger}$}
\address{$^{\dagger}$Department of Physics and Astronomy, University of British
Columbia,\\
6224 Agriculture Road, Vancouver B.C. V6T 1Z1, Canada;}
\address{$^{\dagger\dagger}$Institute for theoretical physics, Utrecht University,\\
Leuvenlaan 4, 3584 CE Utrecht, The Netherlands }
\date{\today}
\maketitle

\begin{abstract}
We study magnetically stabilized nematic order for spin-one bosons
in optical lattices. We show that the Zeeman field-driven quantum phase transitions
between non-nematic Mott states and quantum spin nematic Mott states
in the weak hopping limit are in the universality class of the ferromagnetic XXZ ($S=1/2$)
spin model.
We further discuss these transitions as 
condensation of interacting magnons. 
The development of $O(2)$ nematic order when external fields are applied
corresponds to condensation of magnons, which breaks 
a $U(1)$ symmetry. Microscopically, this results from a coherent superposition
of two non-nematic states at each individual site.   
Nematic order and spin wave excitations around critical points
are studied and critical behaviors are obtained in a dilute gas
approximation.
We also find that spin singlet states are unstable with respect to quadratic Zeeman effects and
Ising nematic order appears in the presence of any finite quadratic Zeeman coupling. 
All discussions are carried out for states in three dimensional bipartite lattices.
\end{abstract}

\begin{multicols}{2}

\narrowtext

\section{Introduction}

When the {\em dimensionless}
exchange coupling strength is strong enough,
Mott states of spinful particles are known to develop certain spin 
order. In the opposite limit,
quantum fluctuations usually restore the broken symmetry
resulting in spin singlet states.
This widely accepted belief however does not exclude, and furthermore implies certain
{\em hidden} fluctuating order in the symmetry restored states.
The preexisting order appears dynamically at certain time and length scales and in general is very
relevant to the low energy physics.
Particularly if an external magnetic field is applied, {\em nontrivial spin (magnetic)
order} might be induced because of the coupling between magnetic 
excitations
and external fields. In other words, an external field can stabilize
a spin order in a parameter regime where order is  
absent in a zero field.

The possible field-induced ordering usually results from condensation of 
magnetic
excitations, or {\em magnons}. The field-induced quantum phase 
transitions between states of different magnetic correlations
and possible
magnetically stabilized order close to critical
points can be investigated by examining magnon excitations
in one of the phases involved in the phase transitions.

A well-known example is the $S=1$ antiferromagnetic
spin-chain\cite{Haldane83,Affleck87,Arovas88,Affleck91,Tsvelik90,Nomura91,Sorensen93}.
The ground state is the spin singlet AKLT state and all spin excitations
are fully gapped by the Haldane gap. An applied Zeeman field along the positive $z$-direction,
although it has zero coupling with the singlet ground state,
couples to spin excitations and lowers the energy of excitations in the $S=1, S_z=1$
branch because of the Zeeman coupling. At a critical field,
the zero momentum excitation becomes degenerate with the spin singlet
ground state signifying a quantum phase transition.

The presence and nature of induced canted Neel order in this case therefore depend {\em
crucially} on interactions
between magnetic excitations in the AKLT phase. One can easily visualize
that the transition is of first order if the interactions of condensed magnons are attractive
or absent. Naturally, the magnetization per lattice site in this case
jumps by a finite value at the critical point as a result of condensation;
furthermore the canted Neel order would not appear in this case. The
phase transition would be simply between a fully polarized state and a spin
singlet state.

However, if interactions between magnons are repulsive,
condensation takes place continuously because of finite chemical potentials
for repulsively interacting magnons.
Thus the magnetization per site,
which is proportional to the density of magnons varies continuously across
the critical point and the transition is of second order.
In this case, the resultant state has canted Neel order.
For the $S=1$ antiferromagnetic spin chain, numerical results show that
the magnetization indeed varies continuously and imply that magnons have
repulsive interactions\cite{Sorensen93}.
In fact, at higher magnetic fields, external fields do induce canted Neel
order in $S=1$ spin chains (only quasi-long-range order prevails in chains).
Condensation of magnons has also been recently studied in three-dimensional
frustrated magnets (see for instance Ref. 
(\onlinecite{Nikuni00,Mishguich04})).

Therefore, to investigate magnetically stabilized order, it is important to
understand interactions between magnons or magnetic
excitations. Generally, microscopic calculations of magnon
interactions are not only very difficult but also practically impossible
for low-spin systems because of uncontrollable approximations involved.
However, in the case we are going to examine we do evaluate the
interactions {\em microscopically} in various situations; therefore we believe
the results about magnetically stabilized quantum spin nematic order and
quantum phase transitions are {\em precise} in this sense.
We also want to emphasize that magnon condensation which interests us in this article
occurs in both high dimensional and one-dimensional {\em non-frustrated} optical lattices;
and the one-dimensional limit will be treated in a separated paper.

The purpose of this article is to understand
magnetically stabilized nematic order of spin-one bosons in optical
lattices. As emphasized above, our starting point will be a series of Mott states with
no nematic order, some of which also have zero coupling with external (linear)
Zeeman fields; our main subject is to investigate
the development of nematic order or
spontaneous symmetry breaking in the $xy$-plane when external magnetic
fields are
applied along the $z$-direction.
In particular, we will focus on the nematic order close to critical points
where our results are actually exact.
Furthermore, the approach we employ here is believed to yield
exact phase boundaries between nematic states and non-nematic Mott states,
a rather remarkable conclusion thanks to a powerful mapping developed
below.

Spin-correlated Mott states for spin-one bosons have recently attracted 
considerable interest\cite{Demler02,Zhou03a,Imambekov03,Snoek04,Zhou03b}.
Theoretical works indicate that spin correlations in Mott states depend on three dimensionless parameters\cite{Zhou03a,Imambekov03}.
The first one is the dimensionless exchange coupling $\eta(=J_{ex}/E_s)$
which is defined as the ratio between the exchange intercation $J_{ex}$ and
the bare spin gap for an individual site $E_s$.
The second parameter is the parity $P$ (even or odd) of the number of
particles per site in Mott states.
The last parameter $D$ is the dimensionality of optical
lattices (we assume all lattices are bipartite).

For $D=2,3$, it was argued that Mott states for all odd $P$
are nematically ordered. In fact irrespective of the exact number of
particles per site, the effective Hamiltonian in the small hopping
limit ($\eta \ll 1$) is equivalent to the bilinear-biquadratic model
for $S=1$ spin chains\cite{Zhou03b}.
For even $P$ on the other hand, nematic and spin singlet Mott states
are present for large $\eta$ and small $\eta$ limits respectively.
The one-dimensional physics ($D=1$) is dominated by quantum fluctuations.
Both dimerized valence bond crystals and non-degenerate spin singlet
states have been found.
As a result of symmetry restoring, the low energy dynamics in Mott states is
mapped into the even- and odd-class quantum dimer models\cite{Zhou03a}.
Furthermore, 
atoms have a tendency to be fractionalized into solitonic {\em elementary} excitations in this 
limit. Superfluid
phases have distinct topological properties and remain to be fully
understood.

Responses of correlated states of spin-one bosons
to external fields are fascinating.
For nematic condensates, the responses
are continuous. The linear coupling between condensates and weak external fields pins the
easy axis
in the $xy$-plane perpendicular to external fields, and $O(3)$ nematic
condensates become $O(2)$ ones, or canted nematic states;
quadratic coupling however, pins the nematic easy axis along the direction
of coupling\cite{Zhou01}.
For spin singlet condensates, the magnetization jumps discontinuously
as a result of a series of level crossings between
states $|S, S_z=S\rangle$ and $|S+2, S_z=S+2\rangle$\cite{Wiemer03}(also see
general discussions about condensates in 
Ref.(\onlinecite{Ho00,Stamper-Kurn98})).
Responses of Mott states to external Zeeman fields and various transitions have
been recently studied in a mean-field approach. Mott states can either
respond to external fields continuously like nematic condensates or
develop interesting magnetization plateaus similar to charge quantization
in a Mott state\cite{Imambekov04}.

In this article, we study magnetically stabilized nematic order in
optical lattices. Particularly we demonstrate the development
of nematic order as {\em repulsively} interacting magnons in non-nematic Mott states condense.
We investigate the induced nematic order associated with the spontaneous breaking
of $O(2)$ nematic symmetry, the magnetization and the spin wave velocity.

In section II, we review the properties of spin singlet Mott states
and introduce a projected nematic order parameter for discussions on
spin partially polarized states.
For a given lattice site with two particles, we show that
nematic order can be established if a spin singlet state
is a superposition with a higher spin
state $|S=2, S_z=2\rangle$. The relative phase between these two states in the
superposition determines the easy
axis of the nematic order parameter, or the orientation of spin nematic states.
In section III, we study the general characterization of nematic order
in spin polarized Mott states; we propose a projected nematic order
parameter which projects away trivial contributions from spin
polarization.

In section IV, we truncate the Hilbert space close to critical magnetic fields
and show that the resultant Hamiltonian is an XXZ ($S=1/2$) pseudo spin
model in an effective
field along the $z$-direction. We carry out microscopic calculations of all
parameters in the effective ferromagnetic XXZ model.
These calculations are done for two particles per site, four particles
per site; in the large $N$ (even) limit; the quantum rotor model
studied in previous works is employed to facilitate calculations.

In section V, close to critical lines and a tri-critical point
we further study the properties of various phases of the XXZ model in both semi-classical
approximation and  dilute gas approximation based on the
Holstein-Primakov boson representation.
We analyze the interactions between Holstein-Primakov bosons or magnons.
We obtain the exact phase boundaries for the ferromagnetic XXZ model by
investigating the instability lines of magnon excitations.
We also discuss the relation between
condensation of magnons close to critical lines, the variation of
magnetization across critical points, and the appearance of
ferromagnetic order.

In section VI, we investigate, in details, the development of magnetically
stabilized nematic order by examining results following the mapping to
the XXZ model and to the Holstein-Primakov condensation problem.

We notice that results about phase boundaries, nematic order and spin
wave velocities in the critical regime can be obtained in a dilute gas
approximation and therefore, remarkably, are exact.
In section VII, we further study the effect of quadratic Zeeman coupling.
Finally, in section VIII, we conclude our investigation on this subject.

\section{Spin Singlet Mott states, Fluctuating Nematic Order and
Projected Nematic Order}

The Hamiltonian for spin-one bosons with antiferromagnetic
interactions in optical lattices in an external field can be conveniently
expressed as\cite{Demler02,Zhou03a,Imambekov03,Snoek04,Imambekov04},

\begin{eqnarray}
&& H=E_s \sum_k {\hat{\bf S}}^2_k +E_c \sum_k {\hat{\rho}}^2_k-\mu_0 \sum_k \hat{\rho_k}
-\sum_{k}\hat{\bf S}_{kz} H_z\nonumber \\
&& -t\sum_{\langle kl\rangle } (\psi^\dagger_{k\alpha}\psi_{l\alpha} +h.c.).
\end{eqnarray}
Here $\psi^\dagger_{k\alpha}(\psi_{k\alpha})$, $\alpha=x,y,z$ are creation
(annihilation) operators for spin-one particles in three different states
at site $k$. $\hat{\bf S}_{k\alpha}=-i\epsilon_{\alpha\beta\gamma}
\psi^\dagger_{k\alpha}\psi_{k\beta}$ and
$\hat{\rho}_k=\psi^\dagger_{k\alpha}\psi_{k\alpha}$ are the spin
and number operators defined
for each lattice site.  It is easy to verify that
$[\hat{S}_{k\alpha},\hat{S}_{k'\beta}]=i\delta_{kk'}$
$\epsilon_{\alpha\beta\gamma}\hat{S}_{k\gamma}$.
$E_{c,s}$ are "bare" charge and spin gaps studied in 
Ref.(\onlinecite{Zhou03a});
$\mu_0$ is the chemical potential.
Finally, the sum over $\langle kl\rangle $ represents the sum over all neigboring sites.

In Mott states, the Hamiltonian can be reduced to the following effective
one,

\begin{eqnarray}
&& H=E_s\sum_k \hat{\bf S}^2_k -\sum_k \hat{\bf S}_{kz} H_z\nonumber \\
&& -J_{ex}\sum_{\langle kl\rangle } (\hat{Q}_{\alpha\beta}(k) \hat{Q}_{\beta\alpha}(l)
+h.c.);\nonumber\\
&& \hat{Q}_{\alpha\beta}(k)=\psi^\dagger_{k\alpha}\psi_{k\beta}-\frac{1}{3}
\delta_{\alpha\beta}\psi^\dagger_{k\gamma}\psi_{k\gamma}.
\label{MottH}
\end{eqnarray}
Eq.(\ref{MottH}) is valid when $t \ll E_c$ and $E_s \ll E_c$; 
$J_{ex}=t^2/2E_c$
is the exchange interaction.

The effective Hamiltonian with zero Zeeman coupling was obtained in a few previous
works; solutions to this Hamiltonian have been studied in various limits.
When the external field is absent and $\eta=J_{ex}/E_s$ is much less than
unity, the ground state is a spin singlet for an even
number of particles per site ($N$).
The spin singlet Mott (SSM) ground state
in this limit is the product of spin singlets
at each individual site (up to a normalization factor)\cite{Zhou03a,Snoek04},

\begin{equation}
\Psi_{SSM}=\prod_k \frac{(\psi^\dagger_{k\alpha}\psi^\dagger_{k\alpha})^{N/2}}
{\sqrt{(N+1)!}}|vac>.
\end{equation}
Meanwhile,
the spin fully polarized (SFP) ground state is the product of on-site maximally
polarized states ,
\begin{equation}
\Psi_{SFP}=\prod_k \frac{(\psi^\dagger_{kx}+i\psi^\dagger_{ky})^{N}}
{\sqrt{ 2^N N!}}|vac>.
\end{equation}

In SSM states, $\langle \hat{Q}_{\alpha\beta}(k)\rangle =0$ as a result of the rotational invariance of the wave 
function and thus there is no nematic
order. The hidden fluctuating nematic order can be studied by examining higher moments. For $N=2$,
one can easily obtain the following results

\begin{eqnarray}
&& \langle \hat{Q}_{\alpha'\beta'}(k)\hat{Q}_{\alpha\beta}(k')\rangle =\nonumber \\
&& \frac{2}{3}\delta_{kk'}(\delta_{\alpha'\beta}\delta_{\beta'\alpha}+\delta_{\alpha' \alpha}
\delta_{\beta\beta'}-\frac{2}{3}\delta_{\alpha'\beta'}\delta_{\alpha\beta}),
\end{eqnarray}
which indicate on-site fluctuating nematic order. More explicitly,
one finds the amplitude of fluctuations of nematic tensor matrix elements
\begin{equation}
\langle \big( \hat{Q}_{\alpha\beta}(k) \big)^2\rangle =\frac{2}{3}(1+\frac{1}{3}\delta_{\alpha\beta}).
\end{equation}

To investigate the responses of spin singlet Mott states or other non-nematic states,
which exhibit certain fluctuating nematic order,
to external fields, it is important to understand how nematic order can be induced by
external Zeeman fields. For this purpose, we focus on the simplest situation where $J_{ex}$ is zero and
treat each site independently. We would like to demonstrate the following important statement:
nematic order appears whenever a spin singlet state and a polarized state are in a linear superposition.

We first consider two particles at one lattice site. The Hilbert space is spanned by five-fold degenerate $S=2$
states and a spin singlet state. When an external field is applied along
the $z$-direction, the five-fold degeneracy
is completely lifted while the maximally polarized state $|S=2, S_z=2\rangle $  approaches the spin singlet ground
state. When the level crossing takes place, the spin projection along $z$-direction jumps by $2\hbar$.
It is obvious that
no nematic order is induced in this simple limit and
there are no transitions between nematic states and spin singlet states.

However,
at the level crossing points, one can further study the properties of coherent
superposition of $|0,0\rangle $ and $|2,2\rangle $ states while the rest
of states are highly excited ones at these crossings.
Let us introduce a coherent state defined in the two-state subspace
as

\begin{eqnarray}
&& |{\bf \Omega}\rangle =\cos\frac{\theta}{2}\exp(-i\frac{\phi}{2})|\! \uparrow \rangle +\sin\frac{\theta}{2}\exp(i\frac{\phi}{2})
|\! \downarrow \rangle ; \nonumber \\
&& |\! \uparrow \rangle =|S=2,S_z=2\rangle 
=\frac{1}{2\sqrt{2}}(\psi^\dagger_x+i\psi^\dagger_y)^2 |vac\rangle ,\nonumber \\
&&|\! \downarrow \rangle =|S=0,S_z=0\rangle =\frac{1}{\sqrt{6}}\psi^\dagger_\alpha\psi^\dagger_\alpha |vac\rangle .
\label{ud}
\end{eqnarray}
Here the unit vector is defined as ${\bf
\Omega}$ $=(\sin\theta\cos\phi,\sin\theta \sin\phi,\cos\theta)$. 
One can easily verify that

\begin{equation}
\langle \hat{S}_z\rangle =2\hbar \cos^2\frac{\theta}{2}.
\end{equation}
$S_z$ reaches the maximum when $\theta=0$ and the minimum when $\theta=\pi/2$.

Direct calculations of the usual nematic order parameter
$Q_{\alpha\beta}$ defined as the expectation value of
the tensor operator $\hat{Q}_{\alpha\beta}$ in Eq.2 suggest that it have a nontrivial
structure in the maximally polarized state.
Namely it contains a) an antisymmetric tensor as a result of spin polarization and b) a traceless
symmetric part which reflects the explicit rotational symmetry breaking by the magnetic field
but is not associated with the
{\em spontaneous} nematic symmetry breaking in a plane perpendicular to the polarization.

To discuss the nematic order in fully or partially polarized states,
it is therefore essential to introduce a projected nematic order
parameter $Q^P_{\alpha\beta}$($\alpha,\beta=x,y,z$),

\begin{equation}
Q^P_{\alpha\beta}=Q_{\alpha\beta}- (Q_{\alpha'\beta'}\Pi^1_{\beta'\alpha'})
\Pi^1_{\alpha\beta}
-(Q_{\alpha'\beta'}\Pi^2_{\beta'\alpha'})\Pi^2_{\alpha\beta}.
\label{PNOP}
\end{equation}
Note that in the projected order parameter, the component associated with the nematic
symmetry breaking remains while
the components associated with polarization have been projected away.
Two tensors we would like to project away are defined as

\begin{eqnarray}
&& \Pi^1_{\alpha\beta}=\frac{1}{\sqrt{2}}\left(\begin{array}{ccc}
0 & -i & 0\\
i & 0& 0\\
0&0&0
\end{array} \right).
\nonumber \\
&& \Pi^2_{\alpha\beta}=\frac{1}{\sqrt{6}}
\left(\begin{array}{ccc}
-1 &0&0\\
0&-1&0\\
0 &0& 2
\end{array}\right)
\end{eqnarray}

One finds that
the nematic order only appears in the $xy$-plane perpendicular to the external fields.
It is indeed straightforward to show that the truncation of $Q^P_{\alpha\beta}$ in the $xy$-plane (i.e.
elements with $\alpha=x,y$ only ) for a coherent state defined above is

\begin{equation}
Q^{Pxy}_{\alpha\beta}=\frac{1}{\sqrt{3}}\sin\theta \left(\begin{array}{ccc}
\cos\phi & \sin\phi\\
\sin\phi & -\cos\phi
\end{array}\right).
\label{NOP1}
\end{equation}

\begin{figure}
\begin{center}
\epsfbox{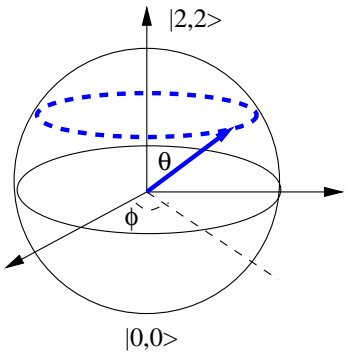}
\leavevmode
\end{center}
\caption{Coherent states $|{\bf \Omega}\rangle $ at the Bloch sphere of pseudo spins.
All states except the north pole $(\theta=0)$ and south pole $(\theta=\pi)$ ones have nonvanishing
expectation value of nematic tensor operator $\hat{Q}_{\alpha\beta}$.}
\end{figure}

Two important features in Eq.(\ref{NOP1}) are worth emphasizing. Firstly,
the two eigenvalues correspond to $\pm \sin\theta/\sqrt{3}$ and are proportional to the
coherence factor in
the linear superposition of coherent states.
They are nonvanishing only if $\theta$ is not zero or $\pi$. Therefore nontrivial nematic
order always appears when $|\! \uparrow \rangle $ and $|\! \downarrow \rangle $ two states are in a linear superposition.

Secondly,
the eigenvector
with the maximal eigen value
represents
the easy axis of nematic order.
And the easy axis is fully characterized by the azimuthal angle of ${\bf \Omega}$. Indeed, one finds that
the easy axis in the $xy$-plane is defined as a 2D unit vector in the $xy$-plane.

\begin{equation}
\omega=(\cos\frac{\phi}{2},\sin\frac{\phi}{2}).
\end{equation}
When $\phi$ varies from $0$ to $2\pi$, the easy axis $\omega$ rotates by $\pi$ angle in the $xy$-plane.
And the nematic order parameter is indeed a tensor constructed out of
the 2D unit vector $\omega$,

\begin{equation}
Q^{Pxy}_{\alpha\beta}=\frac{2}{\sqrt{3}} \sin\theta ({\bf \omega}_\alpha{\bf \omega}_\beta-\frac{1}{2}\delta_{\alpha\beta}).
\end{equation}

\section{Nematic order parameter for spin partially polarized states:\\ General characterization}

As we have already seen in the previous section, the complication of characterizing
nematic order when spins are partially polarized comes from the explicit
symmetry
breaking induced by external fields. So in this case one has to
deal with the tensor $\hat{Q}_{\alpha\beta}$
which has nontrivial elements even without nematic order.
To distinguish the spontaneous symmetry breaking due to the formation of
nematic order from explicit symmetry breaking in the presence of polarization,
special care needs to be taken of the elements which are induced by spin polarization.

A general scheme to project out nematic order parameter tensor appears to be possible, similar to what is
carried out in the previous section. Assume spins are polarized along direction ${\bf s}$ (unit vector).
Introducing two projection tensors

\begin{eqnarray}
\Pi^1_{\alpha\beta}=\frac{1}{\sqrt{2}}
i\epsilon_{\alpha\beta\gamma} {\bf s}_\gamma,
 \Pi^2_{\alpha\beta}=\frac{3}{\sqrt{6}}
({\bf s}_\alpha {\bf s}_\beta -\frac{1}{3}\delta_{\alpha\beta}),
\label{ptensor}
\end{eqnarray}
we again are able to define a projected nematic order parameter as in Eq.(\ref{PNOP}).

When the nematic symmetry is broken along the direction ${\bf{\omega}}$ (unit vector),
in the large $N$ limit one can easily demonstrate that ${\bf s}
\cdot {\bf{\omega}} =0$ following the algebras in 
Ref.(\onlinecite{Zhou01,Zhou03a}); and
${\bf{\omega}}$ and ${\bf s}$ always appear to be
orthogonal. One can further define

\begin{equation}
{\bf m}={\bf s}\times {\bf{\omega}}.
\end{equation}
Then ${\bf{\omega}}$, ${\bf m}$ and ${\bf s}$ form an orthogonal triad.

It is possible to verify the validity of the definition for nematic order parameters given Eqs.(\ref{PNOP}),(\ref{ptensor}).
For instance, one can consider the following spin partially polarized nematic states

\begin{equation}
|\Psi\rangle=\frac{[(\cos\frac{\xi}{2} {\bf n}+i\sin\frac{\xi}{2} {\bf
m})_\alpha\psi^\dagger_\alpha]^N}{\sqrt{N!}}|vac\rangle
\label{gc}
\end{equation}
We have assumed that ${\bf n}$ and ${\bf m}$ are orthogonal, i.e. ${\bf
n} \cdot {\bf m}=0$; $\xi$ varies from $0$ to $\pi$.
Following the discussions in Ref.(\onlinecite{Zhou01,Zhou03a,Snoek04}), states specified in Eq.(\ref{gc}) with $\xi=0$ form a complete set
of $N$-particle-condensate wave functions. And a condensate with total spin $S, S_z$ ($S\leq N$) can be expressed in terms of spherical harmonics $Y_{S,S_z}({\bf 
n})$ in this representation. 

When $\xi\neq 0$,
a state given above is polarized along a direction perpendicular to
${\bf n}$ and ${\bf m}$.
Indeed,

\begin{equation}
\langle \hat{\bf S}\rangle =N \sin  \xi {\bf n}\times {\bf m},
\quad {\bf s}={\bf n}\times {\bf m}.
\end{equation}

A direct calculation shows that the projected nematic order parameter can be expressed in terms of
three orthogonal unit vectors $({\bf n},{\bf m}, {\bf s})$,

\begin{eqnarray}
&& \frac{Q^P_{\alpha\beta}}{N}=\cos^2\frac{\xi}{2} {
n}_\alpha{
n}_{\beta}
+\sin^2\frac{\xi}{2} { m}_\alpha { m}_{\beta}
+\frac{1}{2}{ s}_\alpha{
s}_{\beta}-\frac{1}{2}\delta_{\alpha\beta}.
\end{eqnarray}
Note that the projected nematic order parameter is traceless and fully symmetric\cite{Mueller03}.
One can further truncate the projected matrix in the $({\bf n}, {\bf m})$-plane perpendicular to
${\bf s}$ and indeed find that

\begin{equation}
Q^{P{nm}}_{\alpha\beta}=\frac{N}{2}
\left( \begin{array}{ccc}
\cos\xi & 0 \\
0 & -\cos\xi
\end{array}\right)
\end{equation}
which is diagonal when $\alpha, \beta$ are chosen to be along the axis
${\bf n}$ or ${\bf m}$.

The projected nematic order parameters $Q^P_{\alpha\beta}$, $Q^{Pnm}_{\alpha\beta}$
vanish when spins are fully polarized or $\xi=\pi/2$; and
the nematic matrix has zero eigenvalues.
When $\xi\neq \pi/2$, the matrix has nontrivial eigenvalues
$\pm N/2 \cos\xi$.
The nematic axis ${\bf \omega}$ therefore 
lies along the direction of ${\bf n}$ when $\xi$ varies from $0$ to 
$\pi/2$ and along the direction of ${\bf m}$ when from $\pi/2$ to $\pi$.
The nematic matrix eigenvalues reach maxima when $\xi=0$ or $\pi$, 
representing spin unpolarized
nematic states.

Obviously,
nematic symmetry order can develop along an arbitrary direction 
in a plane perpendicular to ${\bf s}$. In fact,
an $O(2)$ rotation of the orthogonal basis $({\bf n},{\bf m})$  
along 
${\bf s}$ by a $\phi$ angle, while leaves $\langle \hat{\bf S}\rangle $ invariant,
results in a new nematic state with easy axis ${\bf \omega}$; 

\begin{equation}
{\bf{\omega}}=\cos\phi {\bf n}+\sin\phi {\bf m}
\end{equation}
if $\xi \in [0,\pi/2]$,and 

\begin{equation}
{\bf{\omega}}=-\sin\phi {\bf n}+\cos\phi {\bf m}
\end{equation}
if $\xi \in [\pi/2,\pi]$.

In terms of the easy axis ${\bf{\omega}}$, the projected order parameter
can be conveniently expressed as 

\begin{equation}
Q^{Pnm}_{\alpha\beta}=N |\cos\xi | ({\bf \omega}_\alpha{\bf \omega}_\beta
-\frac{1}{2}\delta_{\alpha\beta}).
\end{equation}

To summarize, we have shown that 
a projected traceless nematic tensor order parameter should be introduced to study
nematic ordering in the presence of external fields.

\section{Ferromagnetic XXZ model as the effective Hamiltonian close to
critical points}

\subsection{Phenomenology}

To study the magnetically stabilized nematic order,
we consider a limit when the exchange interaction $J_{ex}$ is much less than $E_s$.
For an even number of particles per site and in the absence of
external fields, the ground state is a spin singlet Mott state
and nematic order is absent. The development of
nematic order first occurs when

\begin{equation}
2  H_z \approx 6 E_s \gg J_{ex}.
\label{lc}
\end{equation}

The Hilbert space for the whole lattice
is a direct product of spin towers ${\cal H}_k$ at each site

\begin{equation}
{\cal H}_{T0}={\cal H}_1 \otimes {\cal H}_2 \otimes {\cal H}_3\otimes
...\otimes {\cal
H}_k\otimes...
\end{equation}
The on-site Hilbert space ${\cal H}_k$ is spanned by
$(N+1)(N+2)/2$ states, with spins equal to $0,2,4,...,N$;
the dimension of the Hilbert space for the whole
lattice is
\begin{equation}
{\cal D}_{T0}=(\frac{(N+1)(N+2)}{2})^{V_T}
\end{equation}
where $V_T$ is the number of lattice sites.

When the external fields satisfying the condition in Eq.(\ref{lc}) are applied
and when $J_{ex}=0$,
at each individual site
the first excited state $|S=2, S_z=2\rangle $ and ground state $|S=0, S_z\rangle $ are
nearly degenerate and are far away from other excited states.
At the point when
the field reaches a value so that
\begin{equation}
 H_z={3E_s},
\end{equation}
level crossing between the the ground state and first excited states occurs
as mentioned briefly in the previous section.
Following discussions in section II, if the hopping or the exchange energy
is set to be precisely zero, then
magnetization jumps and
$Q^P_{\alpha\beta} =0$.
In this case, nematic order is not induced by
external fields.

As shown in the previous section, for nematic order to be present,
two nearly degenerate states have to be in a linear superposition.
In this sense, naturally
it is the exchange process in the vicinity
of level crossing points which eventually leads to nematic order
which doesn't exist in zero fields.
This observation leads us to truncate the on-site Hilbert space
into a two-dimensional one for a pseudo-spin. The truncated Hilbert
space
for the whole lattice is then a product of pseudo spin Hilbert space
${\cal S}_k$
at each site $k$

\begin{equation}
{\cal H}_{Tt}={\cal S}_1\otimes {\cal S}_2\otimes {\cal S}_3...
\otimes{\cal S}_k \otimes...
\end{equation}
and the on-site pseudo spin Hilbert space ${\bf S}_k$ consists of two states:

\begin{equation}
|\! \uparrow \rangle =|S=2, S_z=2\rangle, |\! \downarrow \rangle =|S=0, S_z=0\rangle .
\end{equation}
For two particles per site, the microscopic wave function of these two
states are given in Eq.(\ref{ud}).

The dimension of the truncated space ${\cal D}_{Tt}$ is exponentially small compared with
the
original one ${\cal D}_{T0}$; i.e.,

\begin{equation}
{\cal D}_{Tt}=2^{V_T} \ll {\cal D}_{T0};
\end{equation}
it is also independent of the number of particles per site.
The phenomenology for different even numbers of particles per site
is therefore identical.

The hopping between two nearest neighbors in lattices introduces
exchange interactions between pseudo-spins.
We will present results of microscopic calculations in the following
subsection. Here we provide a phenomonology of this model.
To facilitate discussions, we define $|\! \uparrow \rangle $ and $|\! \downarrow \rangle $ to be two
eigen states of Pauli-matrix $\sigma_z$,

\begin{equation}
\sigma_z |\! \uparrow \rangle =|\! \uparrow \rangle , \sigma_z |\! \downarrow \rangle =-|\! \downarrow \rangle .
\label{Pauli}
\end{equation}
Note that these two pseudo spins are also eigen states of
the spin operator $\hat{\cal S}_z$. Therefore the pseudo spin algebra
corresponds to the projection of usual $SU(2)$ spin algebra in the truncated
pseudo spin space. For instance, one can verify the following
mapping

\begin{equation}
\hbar (\sigma_z+1) \rightarrow \hat{\bf S}_z,
\hbar \sigma^+ \rightarrow \hat {\bf S}^+,
\hbar \sigma^-\rightarrow \hat {\bf S}^-.
\label{Pseudo}
\end{equation}

An important and obvious fact is that single particle hopping conserves
the total spin of two sites and its projection along all
directions including the $z$-direction.
Following Eq.(\ref{Pseudo}), this conservation of spins implies
that any induced exchange coupling have to as well conserve the pseudo spin
defined along $\sigma_z$ axis in the presence of external Zeeman fields.
Furthermore, the superexchange due to virtual hopping between two bosonic
$S=1/2$ pseudo spins results in a ferromagnetic coupling which is to be
further verified by microscopic calculations.

Based on the above considerations, one concludes that the effective
Hamiltonian in the truncated space should be

\begin{eqnarray}
&& \frac{H_{eff}}{J_{ex}}=-2{\epsilon_0} \sum_{\langle kl\rangle }
(\sigma_{k}^+\sigma_{l}^- +\sigma_{k}^-\sigma_{l}^+) \nonumber \\
&& -(\beta+1)\epsilon_0\sum_{\langle kl\rangle }\sigma_{kz}\sigma_{lz}-\epsilon_0 h_z
\sum_k \sigma_{kz}.
\end{eqnarray}
Here $\epsilon_{0},\beta$ depends on microscopic details of states and should
be a
function of the number of particles per
site and $\eta_1$, the ratio between $E_s$ and $E_c$. $h_z$ further
depends on $\eta$ (the ratio between $J_{ex}$ and $E_s$) and the ratio between
external fields $H_z$ and $J_{ex}$.

One can easily recast the Hamiltonian into the following ferromagnetic
$XXZ$ model in an effective external field along the $z$-direction,

\begin{equation}
\frac{H_{XXZ}}{\epsilon_0 J_{ex}}=- \sum_{\langle kl\rangle }
\sigma_{k\alpha}\sigma_{l\alpha}
-\beta\sum_{\langle kl\rangle }\sigma_{kz}\sigma_{lz}-h_z
\sum_k \sigma_{kz}.
\label{XXZ1}
\end{equation}
Because external magnetic fields are applied along the $z$-direction,
with the particular choices of eigen states for the pseudo spin Pauli
matrix $\sigma_z$ in Eq.(\ref{Pauli}),
the Hamiltonian in Eq.(\ref{XXZ1}) also has an $O(2)$ invariance in the 
$xy$-plane.
This $O(2)$ symmetry represents the $O(2)$ nematic symmetry we are going to
examine. The relation between the symmetries of the pseudo spin model and
the microscopic model for spin-one bosons has
been addressed in previous sections.

In general, the truncation can be applied in the vicinities of all
critical
points where level crossings between $|S, S_z=S\rangle $ and $|S+2, S_z=S+2\rangle $ occur,
$S+2 \leq N$. One arrives at the same phenomenology as for the level crossing between the
first two states. Of course $\epsilon_0, \beta$ and $h_z$ then depend on
the states involved in level crossings
and are functions of $S$, $S=0,2,4,...N-2$.
In the next few subsections we are going to calculate $\epsilon_0,\beta$ and $h_z$.

\subsection{Calculations of parameters $\epsilon_0$,$\beta$ and $h_z$ in
the $XXZ$ model}

Microscopic calculations of $\beta$ and $h_z$ though straightforward are
pretty involved. We present results in a few limits; detailed
calculations can be found in appendix A.

\subsubsection{Two particles per site}
There is only one level crossing in this case.
One can verify that

\begin{equation}
\langle \uparrow \!|\hat{Q}_{\alpha\beta}|\! \uparrow \rangle =\left(
\begin{array}{ccc}
\frac{1}{{3}} & {i} & 0 \\
-{i} & \frac{1}{{3}} & 0 \\
0 & 0&-\frac{2}{3}
\end{array}\right),
\end{equation}

\begin{equation}
\langle \downarrow \!|\hat{Q}_{\alpha\beta}|\! \uparrow \rangle =\left(
\begin{array}{ccc}
\frac{1}{\sqrt{3}} & \frac{i}{\sqrt{3}} & 0 \\
\frac{i}{\sqrt{3}} & -\frac{1}{\sqrt{3}} & 0 \\
0 & 0&0
\end{array}\right),
\end{equation}
and

\begin{equation}
\langle \downarrow \! |\hat{Q}_{\beta\alpha}|\! \uparrow \rangle  = \langle \uparrow \!|\hat{Q}_{\alpha\beta}|\! \downarrow \rangle ^{\dagger},
\langle \downarrow \!|\hat{Q}_{\beta\alpha}|\! \downarrow \rangle =0.
\end{equation}

Using the Hamiltonian in Eq.(\ref{MottH}) and taking into account of these 
matrix elements
of $\hat{Q}_{\alpha\beta}$ in the truncated Hilbert space, one
obtains the results for $\epsilon_0,\beta$ and $h_z$. In this particular
case, one finds
$\epsilon_0=4/3$, $\beta=0$ which implies an $O(3)$ symmetry when
the effective
field $h_z$ vanishes (but with a finite external Zeeman field $H_z$). It leads to
a symmetry higher
than the $O(2)$ one in the original problem in the presence of Zeeman field $H_z$.

We believe that the $O(3)$ symmetry found for 2-particles per site is
{\em accidental} and can be removed by taking into account contributions
of order of
$\eta_1=E_s/E_c$.
(see Appendix A for details). The final result can be summarized in the
following equation,

\begin{eqnarray}
&& \epsilon_0=\frac{4}{3} \left( 1+ \frac{E_s}{E_c} \right),
\beta=-\frac{3 E_s}{E_c+E_s}, \nonumber \\
&& h_z = - \frac{9 E_s - 3 H_z - 8 J_{ex} - 2 \frac{E_s}{E_c} J_{ex}}{4 J_{ex} \left(1+ \frac{E_s}{E_c} \right)}.
\end{eqnarray}

\subsubsection{Four particles per site}

Close to level crossing between $|0,0\rangle $ and $|2,2\rangle $,
we find that

\begin{equation}
\langle \uparrow \!|\hat{Q}_{\alpha\beta}|\! \uparrow \rangle =\left(
\begin{array}{ccc}
\frac{11}{22} & i & 0 \\
-i & \frac{11}{22} & 0 \\
0 & 0& -\frac{22}{21}
\end{array}\right),
\end{equation}

\begin{equation}
\langle \downarrow \!|\hat{Q}_{\alpha\beta}|\! \uparrow \rangle =\left(
\begin{array}{ccc}
\sqrt{\frac{14}{15}} & i\sqrt{\frac{14}{15}} & 0 \\
i\sqrt{\frac{14}{15}} & -\sqrt{\frac{14}{15}} & 0 \\
0 & 0&0
\end{array}\right).
\end{equation}

The corresponding parameters $\epsilon_0,\beta$ and $h_z$ are

\begin{equation}
\epsilon_0=\frac{56}{15},
\beta=-\frac{351}{686},
h_z=-\frac{15}{56}(\frac{3E_s-H_z}{J_{ex}}-\frac{536}{147}).
\end{equation}
The effective XXZ model has the desired $O(2)$ symmetry in the plane perpendicular
to the external field.
For four particles, level crossing also happens between $|2,2\rangle $ and $|4, 4\rangle $ states.
Similar calculations have been carried out and presented in Appendix A.

\subsubsection{Large-$N$ limit (even $N$)}

In the large-$N$ limit, one can describe the collective ground state and excited states
in terms of spherical harmonics in a quantum rotor representation. So the spin
singlet ground state and polarized $|S=2, S_z=2\rangle $ wave functions are

\begin{equation}
|\! \uparrow \rangle =\frac{1}{4}\sqrt{\frac{15}{2\pi}}\sin^2 \theta\exp(i2\phi),
|\! \downarrow \rangle =\frac{1}{\sqrt{4\pi}}.
\end{equation}
In the quantum rotor representation, the Hamiltonian is\cite{Zhou01,Demler02,Zhou03a,Snoek04}

\begin{eqnarray}
&& H=E_s \sum_k {\bf S}^2_k -
{H}_z \sum {\bf S}_z
\nonumber \\
&& -J_{ex} \sum_{\langle kl\rangle }[Q_{\alpha\beta}({\bf n}_k)
Q_{\beta\alpha}({\bf n}_l)+h.c.]
\end{eqnarray}
where  ${\bf S}=i{\bf n}\times \partial/\partial {\bf n}$, the spin
operator is defined as the angular momentum of $O(3)$ quantum rotor.
It is a conjugate variable to director ${\bf n}$,

\begin{equation}
[{\bf S}_\alpha, {\bf n}_\beta]=-i\epsilon_{\alpha\beta\gamma}
{\bf n}_\gamma.
\end{equation}

Again the matrix elements of $\hat{Q}_{\alpha\beta}$ are calculated below

\begin{equation}
\langle \uparrow \!|\hat{Q}_{\alpha\beta}|\! \uparrow \rangle =\frac{1}{21}\left(
\begin{array}{ccc}
2 & 0 & 0 \\
0 & 2 & 0 \\
0 & 0& -4
\end{array}\right), \label{upuplargen}
\end{equation}

\begin{equation}
\langle \downarrow \!|\hat{Q}_{\alpha\beta}|\! \uparrow \rangle =\frac{1}{\sqrt{30}}\left(
\begin{array}{ccc}
1 & i & 0 \\
i & -1 & 0 \\
0 & 0&0
\end{array}\right).
\end{equation}

The matrix element $\hat{Q}_{xy}$ vanishes in Eq.(\ref{upuplargen}) as an
artifact of the large-$N$ approximation.
One then obtains all parameters for the XXZ effective model,

\begin{equation}
\epsilon_0=\frac{2}{15},\beta=-\frac{39}{49},
h_z=-\frac{15}{2} (\frac{3E_s-H_z}{J_{ex}}-\frac{8}{147}).
\end{equation}

It is possible to generalize this analysis to level crossing between high
spin states $|S, S\rangle $ and $|S+2, S+2\rangle $ ($S\leq N-2$) and results are qualitative
the same (see Appendix B). In all cases, $\epsilon_{0}$ is positive and $\beta$ is
negative.
In the next section, we are going to examine the consequency of this model.
Particularly we investigate the implications
on magnetically stabilized nematic order and physics around critical
points.

\section{Phase Boundaries of the XXZ Model and
Holstein-Primakov Bosons}

\subsection{Phases of XXZ model}

The general phase diagram in the $(\beta,h_z)$-plane can be easily obtained
first in a mean field
approximation. Later on we argue that the phase boundaries and
solutions obtained in this way in some part of the plane
are actually exact (see Fig. 2).
In the mean field approximation, we introduce $\bf s$ as a
unit vector order parameter
which defines the orientation of spin,
\begin{equation}
\langle {\bf \sigma}\rangle =2S {\bf s};
\end{equation}
here $S=1/2$ is the pseudo spin.

The $\bf s$-dependence of the total
energy comes entirely from
the terms proportional to $\beta$ or $h_z$; that is

\begin{equation}
\frac{E}{\epsilon_0 J_{ex}V_T}=const- 4 d \beta S^2 {\bf s}^2_z
- 2 h_z S {\bf s}_z.
\label{MF}
\end{equation}
where $d=3$ is the dimension of three dimensional cubic lattices. ${\bf s}_z$ varies from $-1$ and $1$.
Minimizing the energy with respect to $\bf s$ one obtains
mean field solutions for various ground states.

Following Eq.(\ref{MF}), when $h_z > -2 d \beta$ and $h_z >0$,
the mean field solution is

\begin{equation}
{\bf s}=(0,0,1)
\end{equation}
representing a fully polarized state
which we call Up-Polarized (UP) phase.

When $h_z< 2 d \beta$ and $h_z <0$,
the mean field solution is

\begin{equation}
{\bf s}=(0,0,-1)
\end{equation}
representing another fully polarized state which we call
Down-Polarized (DP) phases.

In addition, when $-2 d \beta > h_z > 2 d\beta$ and $\beta <0$,
the mean field solution is
\begin{equation}
{\bf s}=(\sin\Theta\cos\Phi,\sin\Theta\sin\Phi,
\cos\Theta), \cos\Theta =-\frac{h_z}{2 d \beta}.
\end{equation}
representing
a ferromagnetically ordered (FO) phase which breaks the in-plane $O(2)$
symmetry spontaneously.
Solutions of different angle $\Phi$ are degenerate and the vacuum
manifold is a unit circle $S^1$.
$\Theta$ varies from $\pi$ to $0$ when $h_z$ increases from $2 d\beta$ to
$-2 d\beta$.

Three phases are separated by a first order transition line
along the $h_z=0$ axis which starts at point $(0,0)$ and
ends at $(\infty, 0)$,
and two other second order phase transition lines (see Fig.2).
These two lines are defined by

\begin{equation}
2d \beta\pm h_z=0;
\end{equation}
both terminate at point $(0,0)$.
Finally $(0,0)$ is a tri-critical point.

Along the first order transition line,
the UP and DP states become degenerate and the ground state breaks
$Z_2$ or Ising type
of symmetry spontaneously. At the tricritical point $(0,0)$, the XXZ model is $O(3)$
rotation invariant and the ground state breaks $O(3)$ symmetry
spontaneously.
At this point, UP, DP and FO states are all degenerate.

In UP and DP phases, the microscopic wave functions for ground states
are, respectively,

\begin{eqnarray}
&& |g_{\uparrow}\rangle =\prod_k |\! \uparrow \rangle _k,
|g_\downarrow \rangle =\prod_k |\! \downarrow \rangle _k; \nonumber \\
&&\sigma_{kz}|\! \uparrow \rangle _k =|\! \uparrow \rangle _k, \sigma_{kz}|\! \downarrow \rangle _k=-|\! \downarrow \rangle _k.
\end{eqnarray}
In the $O(2)$ ferromagnetic phase,

\begin{eqnarray}
&& |g_F\rangle =\prod_k |\Omega\rangle _k, \nonumber \\
&& |\Omega\rangle _k=\cos\frac{\Theta}{2}\exp(-i\Phi)|\! \uparrow \rangle _k+\sin\frac{\Theta}{2}|\! \downarrow \rangle _k;
\label{ferro}
\end{eqnarray}
and $\Theta$ is a function of $\beta, h_z$,
$\cos\Theta=\frac{h_z}{2 d \beta}$, $|h_z| < 2 d |\beta|$.
Solutions in Eq.(\ref{ferro}) are degenerate in the $S^1$-manifold
where $\exp(i\Phi)$ lives and
represent spontaneous $O(2)$-symmetry breaking states.

By examing the microscopic wave functions of UP and DP states,
we notice that
the UP and DP states are non-degenerate exact eigenstates of the pseudo
spin operator $\Sigma_z=\sum_k \sigma_{kz}$.
Meanwhile, the total pseudo spin
projected along the $z$-axis is a conserved quantum number.
So these UP and DP solutions are {\em exact eigenstates} of the XXZ-Hamiltonian.
In the next subsection we are going to show that they are actually
exact ground states when $\beta$ is positive; furthermore we
argue that UP or DP states are also exact ground states
even when $\beta$ is negative and $h_z >- 2 d \beta$
or $h_z < 2 d \beta$.

\subsection{UP and DP states as exact ground states}

When $\beta, h_z$ are both positive,
the UP state presented above is a ground state of both the $O(3)$
isotropic term in the ferromagnetic XXZ model and the terms involving $\beta, h_z$.
So naturally the UP state is the exact ground state of the XXZ model in this limit.
Similarly when $\beta$ is positive and $h_z$ is negative,
the DP state is the exact ground state.

When $\beta <0$ but outside the triangular defined
by the two critical lines $2 d \beta\pm h_z=0$, we are not able to prove
rigorously
that eigenstates in Eq.(\ref{ferro}) are exact ground states.
However, we would like to show that they are locally stable and therefore
we argue that they are likely to be the
exact ground states.

To carry out this part of discussions, we study
the magnon excitation spectrum in
UP and DP phases and show that one-particle magnon excitations
are also exact eigen states; furthermore they are fully gapped
except along the second order transition lines.
The most straightforward approach to study
these excitations is to use the Holstein-Primakov boson
representation for the XXZ model.

In the Holstein-Primakov representation, all spin operators are expressed
in terms of Holstein-Primakov bosons,

\begin{eqnarray}
\sigma^+ &=& \left(\sqrt{ 2 S - c^\dagger c}\right) c \\
\sigma^- &=& c^\dagger \sqrt{ 2 S - c^\dagger c} \\
\sigma_z &=& 2( S- c^\dagger c \label{sigmaz})
\end{eqnarray}
$c^{\dagger}(c)$ is the creation (annihilation) operator of bosons satisfying
the usual bosonic commutation relations
$\lbrack c, c^\dagger \rbrack = 1$;
and the raising and lowering operators are defined as
\begin{equation}
\sigma^+= \frac{\sigma_x + i\sigma_y}{2},
\sigma^-=\frac{\sigma_x - i\sigma_y}{2}.
\end{equation}
One can furthermore verify that
\begin{equation}
\lbrack \sigma_\alpha , \sigma_\beta \rbrack = i 2 \epsilon_{\alpha \beta
\gamma} \sigma^\gamma,
{\bf \sigma} \cdot {\bf \sigma} =4 S(S+1)
\end{equation}

The Hamiltonian of the XXZ model then transforms into:
\begin{eqnarray}
\frac{{H}_{XXZ}}{\epsilon_0 J_{ex}} &=&
-{2} \sum_{\langle  kl \rangle} \sqrt{\lbrack 2 S - c_k^\dagger c_k \rbrack} c_k
c_l^\dagger \sqrt{2 S - c_l^\dagger c_l} \nonumber \\
&&
-{2} \sum_{\langle  kl \rangle} c_k^\dagger \sqrt{(2 S - c_k^\dagger c_k)(2 S - 
c_l^\dagger c_l)} 
c_l 
\nonumber\\
&& - 4 (1 + \beta) \sum_{\langle  kl \rangle } c_k^\dagger c_k c_l^\dagger c_l \nonumber 
\\
&& + 2 (h_z+ 4S(1+\beta)d ) \sum_{k} c_k^\dagger c_k.
\label{HPh}
\end{eqnarray}
In deriving Eq.(\ref{HPh}),
we have neglected a constant term $-4 (1+\beta) S^2- 2 h_z S$ for each lattice site.
In a  semiclassical approximation, one indeed recovers the results
obtained in the previous
section.
Again $S=1/2$.

\begin{figure}
\begin{center}
\epsfbox{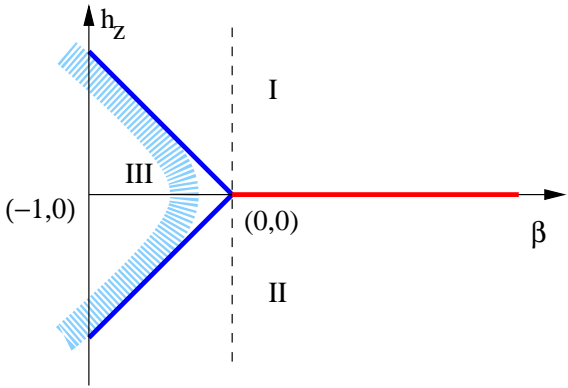}
\leavevmode
\end{center}
\caption{Phases in the ferromagnetic $XXZ$ model.  $\beta$ varies from $-1$ to
$+\infty$.
Region I,II and III represent Up-polarized(UP),Down-Polarized (DP) and Ferromagnetic
Ordered (FO) phases
respectively. Along the blue lines ($-2d\beta \pm h_z=0$), transitions are continuous while along the
red line the transition ($h_z=0,\beta>0$) is a first-order one.
Point $(0,0)$ is the $O(3)$ symmetric tricritical point of the ferromagnetic XXZ model.
As $q$ goes to zero, the interactions between magnons
are repulsive when $\beta<0$ and attractive when $\beta>0$; along the dash line ($\beta=0$),
magnons are non-interacting.
The solutions in the shaded region can be obtained in a dilute gas approximation.}
\end{figure}

To study the excitation spectrum in region I (or II), we first examine the
Hamiltonian in Eq.(\ref{HPh}) in a one-particle subspace next to the exact 
eigen states of UP (or
DP).
First one notices that a UP state is an exact vacuum for Holstein-Primakov bosons;
that is
\begin{equation}
c_k |g_\uparrow \rangle =0,
\sigma_{kz}|g_\uparrow \rangle =|g_\uparrow \rangle .
\end{equation}
for any lattice  $k$.

One-particle excitations we are interested in live in a subspace
of single Holstein-Primakov boson;
that is in a space spanned by states
\begin{equation}
c^{\dagger}_k |g_\uparrow\rangle
\end{equation}
defined at each lattice site $k$.
Since the total number operator of Holstein-Primakov bosons commutes with
the Hamiltonian

\begin{equation}
[N_c, H]=0, N_c=\sum_{k} c_k^{\dagger}c_k,
\end{equation}
$N_c$ is a conserved
quantum number.
We can then diagonalize the Hamiltonian in this one-particle
subspace where
$N_c=1$.

In the subspace,
we obtain the following effective Hamiltonian

\begin{eqnarray}
&& \frac{{H}_{XXZ}^{o.p.}}{\epsilon_0 J_{ex}} =
\sum_{\bf q}
\epsilon_{\bf q}
c_{\bf q}^\dagger c_{\bf q}; \nonumber \\
&& \epsilon_{\bf q}=
8S(d- \sum_{\alpha=x,y,z} \cos{\bf q}_\alpha a )+
2{h_z} + 8 S d \beta.
\label{op}
\end{eqnarray}
The superscript {\em o.p.} stands for the "one particle" subspace.
Eq.(\ref{op}) indicates the dispersion relation of one-particle states.

In particular, it yields a fully gapped magnon spectrum in region I;
the gap vanishes only along the second order transition line where
$\beta <0$ and $h_z =\pm 2 d \beta$.
Especially magnons are fully gapped along the first order phase transition
line $\beta>0$ and $h_z=0$.
When $|h_z| <-2 d\beta$ and $\beta<0$, one-particle states (or magnon
excitations) start to
have lower energies than the vacuum state.
This indicates condensation of Holstein-Primakov
bosons which we are turning to.

To conclude we find that DP and UP states are exact ground
states of the XXZ model in region I and II;
magnon excitations in these phases are fully gapped.
Along the mean field second order transition lines, magnons become
gapless excitations. Further decreasing $h_z$ results in instability
of magnon excitations. So we believe that the transition lines in the mean
field theory represent the exact phase boundaries.

\subsection{Condensation of Interacting Magnons
and Emergence of Ferromagnetic Order in the XXZ model}

As discussed in the introduction, the dynamics of condensation of magnons
depends
crucially on the interactions between magnons. To
study the region close to critical lines
where the condensed particle
density should actually be very low,
we only take into account two-body interactions and
apply a dilute gas expansion. The results
we derive in this subsection are valid in the shaded critical regions
(see Fig.2) where

\begin{equation}
\frac{|h_z \pm 2d \beta|}{|h_z|}\ll 1.
\end{equation}
And as $h_z\pm 2d \beta$ approaches zero, the results become
exact\cite{hd}!
We will present the calculations along the upper transition line
defined by $h_z+2d\beta=0$;
the results are then generalized to the lower transition line
$h_z-2d\beta=0$.

In the dilute gas limit which interests us,
the number of Holstein-Primakov bosons per lattice site is much less than one i.e.
\begin{equation}
n_c=\langle c^{\dagger}_kc_k\rangle  \ll 1.
\label{dilute}
\end{equation}
For this reason, one can expand the nonlinear operators of $\sigma^{\pm}$ in terms of $n_c$;
especially,
\begin{equation}
\sqrt{2S- c^\dagger c} =
\sqrt{2S}\left(1-\frac{c^\dagger c}{4S} +O(n^2_c)... \right)
\end{equation}
for $S=1/2$.
The resultant many-body Hamiltonian up to the second order of $n_c$ is

\begin{eqnarray}
&& \frac{{H}_{XXZ}}{\epsilon_0 J_{ex}} =
\sum_{\bf q}
\epsilon_{\bf q}
c_{\bf q}^\dagger
c_{\bf q} \nonumber \\
&& -4 d \beta \sum_{{\bf q}_1,{\bf q}_2,{\bf q}_3}
c_{{\bf q}_1+{\bf
q}_3}^\dagger
c_{{\bf q}_2-{\bf q}_3}^\dagger
c_{{\bf q}_1} c_{{\bf q}_2}.
\label{int}
\end{eqnarray}
The first term is identical to $H^{o.p.}_{XXZ}$, the exact Hamiltonian projected in the one-particle subspace;
the second term describes magnon-magnon interactions.
This Hamiltonian is applicable in a dilute limit where Eq.(\ref{dilute}) 
is satisfied.
The sign of interaction at $q=0$ or small $q$ limit is determined by
$\beta$.
When $\beta$ is positive, magnon interactions are attractive
and when negative magnon interactions are positive.

When magnons are ideal ($\beta= 0$),
all magnons condense when the energy gap in the spectrum closes at $h_z=0$.
This leads to an abrupt jump in magnetization which corresponds
to the field-driven first order phase transition from UP to DP phase at the tricritical point $(0,0)$ (along the dashed line).
And in this case external fields do not induce nematic order.
This is consistent with mean field results discussed in the previous
subsection. One can in principle generalize this argument to the case when $\beta <0$ and arrive at
similar conclusions.

When magnons' interactions are positive, 
following Eq.(\ref{int}) the chemical potential of magnons in the dilute 
gas limit (differing from $\mu_0$ of
atoms) is

\begin{equation}
\mu=-8 n_0 d \beta +O(n_0^2)
\label{chem}
\end{equation}
where $n_0$ is the 
number of magnons
per lattice site. This is similar to weakly interacting gases of bosons in continuum 
limit\cite{Nozieres}. The chemical potential defined in this way only depends on
intrinsic parameters $\beta$ which have been evaluated microscopically and
is independent of external Zeeman fields.  
The energy of the magnon condensate per lattice site is therefore

\begin{equation}
\frac{E(n_0)}{\epsilon_0 J_{ex} V_T}=2 (|h_z| + 2 d \beta) n_0 - 4 n_0^2 d \beta.
\end{equation}
Minimizing the energy with respect to $n_0$ yields 

\begin{equation}
n_0=\frac{1}{2}[ 1+\frac{|h_z|}{2 d\beta}].
\end{equation}
which is a continuous function of $h_z$.
$n_0$ is much less than one in the critical region where 
Eq.(\ref{dilute}) is 
satisfied. At the transition point $|h_z|=-2d \beta$,
the magnon density per lattice site either vanishes or is equal to one, i.e. $n_0=0,1$.
Furthermore, if one extrapolates to the $h_z=0$ case, one obtains $n_0=1/2$,
that is half magnon per lattice site. 
$\sigma_z=0$ as expected.

Note that
the ground state in this case is not the vacuum of Holstein-Primakov bosons
but instead the vacuum defined by Bogolubov quasi-particles.
The Bogolubov
excitations are created by the following operators

\begin{eqnarray}
&& \gamma^{\dagger}_{\bf q}=u({\bf q}) c^{\dagger}_{\bf q} + v({\bf q}) c_{-\bf q};\nonumber\\
&& u^2({\bf q})=\frac{1}{2}(1+\frac{\epsilon_{\bf q}+\mu}{\sqrt{\epsilon_{\bf q}^2+2\epsilon_{\bf q} \mu}}),
\nonumber\\
&& v^2({\bf q})=-\frac{1}{2}(1-\frac{\epsilon_{\bf q}+ \mu}{\sqrt{\epsilon_{\bf q}^2+2\epsilon_{\bf q} \mu}})
\label{Bogo}
\end{eqnarray}
where $\mu$ is the chemical potential of magnons defined before and
the kinetic energy
$\epsilon_{\bf q}= 2 |{\bf q}|^2a^2$ is written in a dimensionless unit.
The dispersion of quasi-particles is phonon-like $\omega_q=v_s |{\bf q}|$ at small energies;
taking into account the chemical potential in Eq.(\ref{chem}), we obtain
\begin{equation}
v_s=v_{s0}\sqrt{1+\frac{|h_z|}{2d\beta}},
v_{s0}=4a\sqrt{-d\beta}.
\end{equation}
This agrees with the semiclassical solutions obtained in Eq.(\ref{sm}) in 
Appendix C.

These results indicate that the
physics in the XXZ model close to the second order critical lines is
indeed equivalent to
interacting dilute magnons defined by Holstein-Primakov bosons;
especially the emergence of ferromagnetic ordering occurs when condensation
of magnons takes place.  As in the usual c-number approximation for condensed
bosons, we approximate

\begin{equation}
c^\dagger_{q=0}=c_{q=0}=\sqrt{n_0 V_T} \exp(i\Phi).
\end{equation}
Substituting this result into the expressions for $\sigma^{\pm},\sigma_z$
in Eq.(\ref{sigmaz}), we obtain

\begin{eqnarray}
&& \langle \sigma_x\rangle =2\sqrt{2S}\sqrt{n_0} \cos\Phi, \nonumber \\
&& \langle \sigma_y\rangle =2\sqrt{2S}\sqrt{n_0}\sin\Phi, \nonumber \\
&& \langle \sigma_z\rangle =2S-2n_0.
\end{eqnarray}
And again $S=1/2$.

Correspondingly,
the Bogolubov quasi-particles represent the spin wave excitations in $O(2)$
ferromagnets. Following Eqs. \ref{sigmaz}, \ref{Bogo}, one can express $\sigma_{x,y}$
in term of $\gamma^{\dagger}$ and  $\gamma$,

\begin{eqnarray}
&& \delta \sigma_x({\bf r})=
\sum_{\bf q \neq 0} \exp(i{\bf q}\cdot {\bf r})
\frac{\sqrt{2S}}{2
\sqrt{V_T}}(u({\bf q})-v({\bf q}))(\gamma^{\dagger}_{\bf q}+\gamma_{-{\bf q}}),\nonumber\\
&& \delta \sigma_y({\bf r})=\sum_{\bf q\neq 0} \exp(i{\bf q}\cdot {\bf r})
\frac{\sqrt{2S}}{2i
\sqrt{V_T}}(u({\bf q})+v({\bf q}))(\gamma^{\dagger}_{\bf q}-\gamma_{-{\bf q}}).\nonumber\\
\end{eqnarray}
Consider a single quasi-particle state

\begin{equation}
|{\bf q}_0\rangle =\gamma^{\dagger}_{{\bf q}_0}|vac\rangle .
\end{equation}
Spin correlations in this single particle state are

\begin{eqnarray}
&& \langle (\delta \sigma_x({\bf r})-\delta\sigma_x(0))^2\rangle
=\frac{2S}{{V_T}}(u({\bf q}_0)-v({\bf q}_0))^2  \sin^2\frac{{\bf q}_0\cdot \bf r}{2},
\nonumber\\
&& \langle (\delta \sigma_y({\bf r})-\delta\sigma_y(0))^2\rangle
=\frac{2S}{{V_T}}(u({\bf q}_0)+v({\bf q}_0))^2
\sin^2\frac{\bf q_0\cdot \bf r}{2},
\nonumber\\
&& \langle (\delta \sigma_x({\bf r})\delta\sigma_y(0))\rangle
=-\frac{S}{V_T} \sin{\bf q}_0\cdot {\bf r}.
\end{eqnarray}

Remarkably, the corresponding orientation of pseudo spin ${\bf s}$
derived
in the dilute gas approximation is
precisely the same as the semi-classical results obtained in subsection A;
close to the critical line, we notice that

\begin{equation}
\cos\Theta=1-2n_0, \sin\Theta=2\sqrt{n_0}.
\end{equation}
In the next section we are going to discuss the implications of the mapping on
correlated atoms, especially magnetically stabilized nematic order.
Since the semiclassical solutions turn out to be exact along the critical lines,
we would like to believe that solutions are also valid in the
ferromagnetic ordered phase, at least qualitatively and can be extrapolated deep into that phase.

\section{Nematic Order and Phase Boundaries
of Magnetically Stabilized nematic Mott States}

Let us turn to the problem of Mott states of spin-one bosons.
Following discussions in section II, one finds that DP states
correspond to spin singlet Mott (SSMI) states and UP states to
spin fully polarized Mott (SFPMI) states.

The FO states breaking the $O(2)$ symmetry represent quantum spin nematic states with
easy axis determined by the projection of pseudo spin orientation ${\bf
s}$ in the $xy$-plane.
Indeed, for two particles per site the wave function of FO states in 
Eq.(\ref{ferro})
indicates the following spin
correlated Mott states for spin one bosons,

\begin{eqnarray}
&& \Psi_{NM}=\prod_{k} [\cos\frac{\Theta}{2}\exp(-i\frac{\Phi}{2})
\frac{(\psi^\dagger_{kx} + i\psi_{ky}^\dagger)^2}{2\sqrt{2}}\nonumber \\
&&+\sin\frac{\Theta}{2}\exp(i\frac{\Phi}{2})
\frac{\psi^\dagger_{k\alpha}\psi^\dagger_{k\alpha}}{\sqrt{6}}]|vac\rangle .
\end{eqnarray}
$\Theta$ is a function of $\beta$ and $h_z$ as given in section V.A,
\begin{equation}
\cos\frac{\Theta}{2}=\sqrt{\frac{1}{2}-\frac{h_z}{4d\beta}},
\sin\frac{\Theta}{2}=
\sqrt{\frac{1}{2}+\frac{h_z}{4d\beta}};
\end{equation}
As $h_z$ varies from $-2 d\beta$ to $2d \beta$,
$\Theta$ varies from $0$ to $\pi$.
And $\Phi \in [0,2\pi]$ represents an $S^1$-manifold for the spontaneous
symmetry breaking solutions.

For cold atoms, our calculations show that in all cases $\beta$ is
negative. When magnetic fields are varied, the trajectory in the
$\beta-h_z$ planes (see Fig.2.) moves vertically at a given negative $\beta$.
And as magnons are repulsive, magnetic fields stabilize nematic order
via the continuous process of condensation of magnons. So as the magnetic field
increases, phases encountered are spin singlet Mott states,
nematic Mott states (partially polarized) and spin fully polarized states.
Here we will focus on the nematic state.

\begin{figure}
\begin{center}
\epsfbox{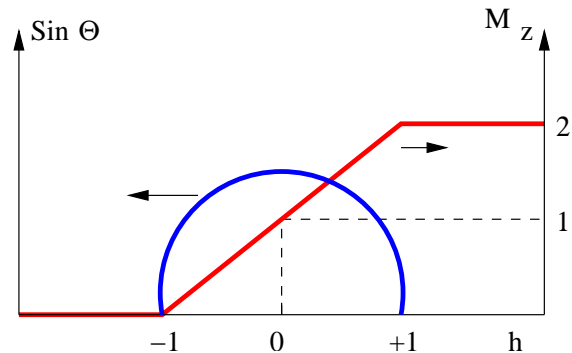}
\leavevmode
\end{center}
\caption{Magnetization $M_z$(in units of $\hbar$) and $\sin\Theta$ as a function of
$h=h_z/2 d\beta$.  $\sin\Theta$ defined in Eqs.(51),(81) is proportional to the nematic order.
Results around upper or lower critical points are obtained in
a dilute gas approximation and are exact. Notice that the nematic order $\sin\Theta$ reaches the maximum at zero $h_z$
where $M_z$ is equal to $\hbar$.}
\end{figure}

The projected nematic order parameter for
the constructed nematic Mott state is given in Eq.(\ref{NOP1}) with
$\phi=\Phi$ and
$\theta=\Theta$:

\begin{eqnarray}
&& Q^{Pxy}_{\alpha\beta}=\frac{1}{\sqrt{3}}
\sqrt{1-\frac{h_z^2}{4d^2 \beta^2}}
({\bf \omega}_\alpha{\bf \omega}_\beta-\frac{1}{2}\delta_{\alpha\beta}),
\nonumber \\
&& {\bf{\omega}}=(\cos\frac{\Phi}{2},\sin\frac{\Phi}{2}).
\end{eqnarray}
The nematic order vanishes along the second order critical lines $|h_z|=-2d\beta$ and reaches maxima
when
level crossing takes place in an isolated lattice site, i.e. at $h_z=0$.

This spin partially
polarized nematic Mott state (SPPNMI) has spin polarization

\begin{equation}
M_z=\langle {\bf S}_z\rangle =  \hbar (-\frac{h_z}{2 d \beta}+1).
\end{equation}
Spins are fully polarized at one of the critical lines ($M_z=2\hbar$)
when $h_z +2 d\beta=0$ and the spin polarization vanishes ($M_z=0$) at the other
critical line $h_z-2 d\beta=0$. In between, $M_z$ varies
continuously
from $0$ to $2\hbar$ and is precisely equal to $\hbar$ when $h_z$
vanishes and the nematic order reaches the maximum.

Finally,
the phase boundaries for SSMI, SFPMI and SPPNMI can be obtained by
substituting the field dependence of $\beta,h_z$ derived in section IV.B into the expression for
critical lines in the XXZ model,

\begin{equation}
h_z(\frac{E_s}{J_{ex}},\frac{H_z}{J_{ex}})\pm 2 d\beta(\frac{E_s}{E_c})=0.
\end{equation}
This results in critical fields for various $N$. Especially one determines
the upper and lower critical fields ($H^{\pm}_{zc}$) between which nematic order develops for the {\em first} time when
magnetic fields increase from zero.

For $N=2$, the upper and lower critical fields are

\begin{equation}
H^{\pm}_{zc}=3E_s-\frac{8}{3}J_{ex} \pm 24 \frac{E_s}{E_c} J_{ex};
\label{ul1}
\end{equation}
for $N=4$, these fields are

\begin{equation}
H^{\pm}_{zc}=3E_s-\frac{536}{147}J_{ex} \pm  \frac{24\times 117}{245} J_{ex}.
\label{ul2}
\end{equation}

At the large-$N$ limit, one obtains

\begin{equation}
H^{\pm}_{zc}=3E_s-\frac{8}{147}J_{ex} \pm  \frac{156}{245} J_{ex}.
\label{ul3}
\end{equation}
We have set $d=3$ in deriving Eqs.(\ref{ul1}),(\ref{ul2}),(\ref{ul3}).

\section{Effects of quadratic Zeeman coupling}

In this section, we are going to briefly discuss the effect of quadratic Zeeman terms which generally 
are present in
atomic gases\cite{Ho00,Stamper-Kurn98}. This kind of external perturbations only
conserves the spin projection along the direction of
external fields
but does not conserve the total spin of the many-body states under consideration and therefore
has distinctly different effects on spin singlet Mott states.
Namely, such external fields would induce nematic order at any small but finite coupling. In other words,
spin singlet Mott states are unstable with respect to these perturbations.

To demonstrate this phenomenon, we consider spin singlet Mott states in the presence of the following
quadratic Zeeman perturbation\cite{Wiemer03,Zhou01},

\begin{equation}
H_p=-H_{q.z.} \sum_k \hat{Q}_{\alpha\beta}(k) ({\bf n}_\alpha{\bf n}_\beta-\frac{1}{3}\delta_{\alpha\beta});
\end{equation}
here $\hat{Q}_{\alpha\beta}(k)$ is the nematic operator defined at the beginning of section II (Eq.2) and
${\bf n}$ characterizes the orientation of quadratic Zeeman fields which we choose to be along the $z$-direction.
$H_{q.z.}$ is the strength of the quadratic Zeeman coupling.
(This perturbation differs slightly from the one used in 
Ref.(\onlinecite{Wiemer03,Zhou01}) by a singlet
operator which doesn't contribute to the quantity we are calculating here.)

One notices that indeed this quadratic Zeeman term does not communte with the total spin 
operator defined at any individual lattice site, $\hat{\bf S}^2_{k}$; however it does 
communte with the operator 
$\hat{\bf S}_{kz}$,

\begin{eqnarray}
&& [\hat{\bf S}_{kz},
\sum_{k'} \hat{Q}_{\alpha'\beta'}(k') \delta_{\alpha'z}\delta_{\beta'z}]=0,\nonumber \\
&& \hat{\bf S}_{kz}=-i\sum_k 
\epsilon_{z\alpha\beta}\psi^{\dagger}_{k\alpha}\psi_{k\beta}.
\end{eqnarray}
So what it does is to cause transitions between states with different on-site spin quantum numbers ${S}_k$ 
but with identical
spin projection along the $z$-direction, ${\bf S}_{kz}$; thus it does not lead to
transitions between
different ${\bf S}_{kz}$ subspaces.
For a spin singlet Mott state, this perturbation results in transitions
between on-site singlet states ($S_k=0$) and non-singlet states ($S_k \neq 0$) 
in the subspace of ${\bf S}_{kz}=0$.

For two particles per site and in the zero hopping limit, we find that these transitions lead to coherent superposition of
two states in $S_z=0$ subspace: $|0,0\rangle $ state and

\begin{equation}
|2,0\rangle =\frac{1}{2\sqrt{3}}(3\psi^\dagger_z\psi^\dagger_z
-\psi^\dagger_\alpha\psi^\dagger_\alpha)|vac\rangle , \alpha=x,y,z.
\end{equation}
For instance, in the first order perturbation expansion the ground state wave function is

\begin{eqnarray}
&& \delta \Psi = \frac{\sqrt{2} H_{b.z.}}{9 E_s} \sum_k \frac{1}{2\sqrt{3}}(3\psi^\dagger_{kz}\psi^\dagger_{kz}
-\psi^\dagger_{k\eta}\psi^\dagger_{k\eta})
\nonumber \\
&& \otimes \prod_{l\neq k} \frac{1}{\sqrt{6}}\psi^\dagger_{l\eta'}\psi^\dagger_{l\eta'} |vac\rangle .
\end{eqnarray}

A direct calculation shows that the nematic order is induced continuously as the quadratic coupling is applied

\begin{equation}
\langle {\hat Q}_{\alpha\beta}\rangle =\frac{2}{3}\frac{H_{b.z.}}{E_s}
({\bf n}_\alpha{\bf n}_\beta-\frac{1}{3}\delta_{\alpha\beta}).
\end{equation}
This dependence is very different from the linear-Zeeman field dependence of nematic order which exhibits a critical
field below which spin singlet Mott states are stable. As expected,
quadratic Zeeman effects are more effective in stabilizing
spin nematic Mott states. Furthermore, the easy axis of the induced nematic tensor order parameter is
pinned along the direction of external fields, ${\bf n}$ and the resultant states are Ising nematically
ordered instead of $O(2)$ or $O(3)$ nematic states discussed
before.

\section{Conclusions}

To summarize, in this article we have investigated magnetically stabilized fluctuating spin nematic order.
We have shown that nematic order can develop
when two non-nematic states at a lattice site are in a linear
superposition in the presence of external fields.
When external fields are applied, even small superexchange coupling
could lead to such a linear superposition and nematic order emerges even
though no spontaneous symmetry breaking occurs in zero field.

We have also mapped the problem of spin-one bosons with
antiferromagnetic interactions in an external
field to the ferromagnetic XXZ spin ($S=1/2$) model.
We find that the field-driven quantum phase transitions belong to the universality class of the ferromagnetic $XXZ$
model ($S=1/2$).
Spontaneous symmetry breaking in the $xy$-plane in this
effective ferromagnetic XXZ model corresponds to
planar nematic ordering in the underlying atomic states.
In all non-nematic Mott states which interest us,
interactions between
magnons are repulsive. Therefore when the external field reaches a
critical one, condensation and thus phase
transitions are continuous.

We also show that the breaking of the $U(1)$ symmetry in magnon Bose condensates
results
in breaking of the $O(2)$ nematic symmetry in the $xy$-plane perpendicular to
external fields. The Bogoliubov quasi-particles of condensates are
precisely the
spin wave excitations in the $O(2)$ nematic states.
So the nematic order is stabilized when the field exceeds a critical one and magnons condense.
We have also obtained the microscopic wave functions of ordered states and
spin wave excitations.

Finally we find that for a spin singlet Mott state the fluctuating nematic order can be stabilized by any small but finite quadratic Zeeman effects.
Namely, the nematic order parameter varies continuously in the presence of quadratic Zeeman effects.

F.Z. would like to thank P.Wiegmann for useful discussions about fermionization
of magnons; he also would like to thank KITP, UCSB for its hospitality
during the workshop on Quantum gases in May-July, 2004.
M.S. and J.W. are supported by the FOM (Netherlands) under contract 02SIC25 and NWO-MK  "Projectruimte" 00PR1929.
This project is also in part supported by the U.S. National Science Foundation Under Grant No.PHY 99-07949 (F.Z.), 
a research Grant from UBC (F.Z.) and NSERC (F.Z., I.A.) and CIAR (I.A.).

\appendix

\section{Effective XXZ model for various numbers of particles
per site}

\subsection{Microscopic Hamiltonian}

To break the O(3) symmetry in the XXZ model for two-particles, we
keep higher order terms of $o(E_s/E_c)$.
The effective Hamiltonian for spin-one bosons in the Mott state of
an optical lattice in the presence of a magnetic field
(in the $z$-direction) can be derived as:

\begin{eqnarray}
\label{hamil}
{H} &=&  \left(E_s - 6 J_{ex} \frac{E_s}{E_c} \right)
\sum_{k} \hat {\bf S}_k^2
-H_z \sum_k\hat {\bf S}^z_k
\\ &&
- 2 J_{ex}\left( 1+ \frac{E_s}{E_c} \right) \sum_{\langle  kl \rangle} \hat
Q_{k,\alpha \beta} \hat Q_{l, \beta \alpha}
\nonumber \\ &&
+ 2 J_{ex} \frac{E_s}{E_c} \sum_{\langle  kl  \rangle} (\hat {\bf S}_k +
\hat {\bf S}_l)^2
\nonumber \\
&&+ J_{ex} \frac{E_s}{E_c}
 \sum_{\langle  kl \rangle} \text{Tr}
\lbrack {A}_k {B}_l + {A}_l {B}_k \rbrack.
\end{eqnarray}
We have introduced the operators
${B}_{\eta \xi} = \psi_\eta^\dagger \psi_\xi$ and ${A}_{\eta \xi} = \psi_{\eta}^\dagger \psi_{\xi}^\dagger
\psi_{\beta} \psi_{\beta}- \psi_{\alpha}^\dagger \psi_{\alpha}^\dagger \psi_{\eta} \psi_{\xi}$.

\subsection{ $N=2$ case}
For $N=2$ we have only the possibility of looking at the transition between the states
$|\! \uparrow \rangle  = |2,2\rangle$ and
$|\! \downarrow \rangle =|0,0 \rangle$
. The relevant non vanishing matrix-elements are (again
$\langle \uparrow \! | \hat Q_{\alpha \beta} | \! \downarrow \rangle =
\langle  \uparrow \!| \hat Q_{\alpha \beta} | \!\uparrow \rangle^\dagger$):

\begin{eqnarray}
\langle  \uparrow \!| \hat Q_{\alpha \beta} | \!\uparrow \rangle &=&
\left( \begin{array}{ccc} \
\frac{1}{3} & i & 0 \\
- i & \frac{1}{3} & 0 \\
0 & 0 & -\frac{2}{3} \end{array} \right), \\
\langle  \downarrow \!| \hat Q_{\alpha \beta} | \!\uparrow \rangle &=&
\left( \begin{array}{ccc} \
\frac{1}{\sqrt{3}} & \frac{i}{\sqrt{3}} & 0 \\
\frac{i}{\sqrt{3}} & -\frac{1}{\sqrt{3}} & 0 \\
0 & 0 & 0 \end{array} \right).
\end{eqnarray}

\begin{eqnarray}
\langle  \uparrow \! | \mathcal{N} | \! \uparrow \rangle  &=& \frac{1}{\sqrt{3}}
\left( \begin{array}{ccc}
1	&	i	&	0  \\
-i	&	1	&	0  \\
0	&	0	&	0
\end{array} \right), \\
\langle  \uparrow \! | \mathcal{N} | \! \downarrow \rangle  &=& \frac{1}{\sqrt{3}}
\left( \begin{array}{ccc}
1	&	-i	&	0  \\
-i	&	-1	&	0  \\
0	&	0	&	0
\end{array} \right), \\
\langle  \downarrow \! | \mathcal{N} | \! \uparrow \rangle  &=& \frac{1}{\sqrt{3}}
\left( \begin{array}{ccc}
1	&	i	&	0  \\
i	&	-1	&	0  \\
0	&	0	&	0
\end{array} \right), \\
\langle  \downarrow \! | \mathcal{N} | \! \downarrow \rangle  &=& \frac{2}{3} \left( \begin{array}{ccc}
1	&	0	&	0  \\
0	&	1	&	0  \\
0	&	0	&	1
\end{array} \right).
\end{eqnarray}
\begin{eqnarray}
\langle  \uparrow \! | \mathcal{A} | \! \downarrow \rangle  &=& \sqrt{3} \left( \begin{array}{ccc}
1	&	-i	&	0  \\
-i	&	-1	&	0  \\
0	&	0	&	0
\end{array} \right), \\
\langle  \downarrow \! | \mathcal{A} | \! \uparrow \rangle  &=& \sqrt{3} \left( \begin{array}{ccc}
-1	&	-i	&	0  \\
-i	&	1	&	0  \\
0	&	0	&	0
\end{array} \right).
\end{eqnarray}

Using these results we find that the effective Hamiltonian turns out to be:

\begin{eqnarray}
{ H} &=& -J_{ex}
\frac{4}{3} \left(1+ \frac{E_s}{E_c} \right) \sum_{\langle  kl \rangle}
\sigma_k\cdot \sigma_l \\ &&
+ 4 J_{ex} \frac{E_s}{E_c} \sum_{\langle  kl \rangle} \sigma_k^z \sigma_l^z
\nonumber \\ &&
- \left(H_z - 3 E_s + \frac{8}{3} J_{ex} + \frac{2}{3} \frac{E_s}{E_c} J_{ex}
 \right) \sum_k \sigma_k^z. \nonumber
\end{eqnarray}
This is the Hamiltonian for the XXZ-model:
\begin{equation}
\frac{{H}_{XXZ}}{\epsilon_0 J_{ex}} = - \sum_{\langle   kl
\rangle} \sigma_k \cdot \sigma_l
- \beta \sum_{\langle   kl \rangle}
\sigma_k^z \sigma^z_l - h_z \sum_k \sigma_k^z
\end{equation}
with
\begin{eqnarray}
\epsilon_0 &=&  \frac{4}{3} \left(1+ \frac{E_s}{E_c} \right) \\
\beta &=& - \frac{3 E_s}{E_c + E_s} \\
h_z &=& - \frac{9 E_s - 3 H_z - 8 J_{ex} - 2 \frac{E_s}{E_c} J_{ex}}{4 J_{ex} \left(1+ \frac{E_s}{E_c} \right)}.
\end{eqnarray}

\subsection{$N=4$ case}
For four particles per site, there are two possible transitions:
$|0,0\rangle \rightarrow |2,2\rangle \rightarrow |4,4 \rangle$. We
will consider both transitions. In both cases
the correction of order $J_{ex} \frac{E_s}{E_c}$
turns out to be not particularly interesting.

\subsubsection{$|0,0 \rangle \rightarrow |2,2\rangle$}
We define again $|\! \uparrow \rangle  = |2,2\rangle$ and $|\! \downarrow \rangle =|0,0 \rangle$. The
relevant non vanishing matrix-elements are:
\begin{eqnarray}
\langle  \uparrow \!| \hat Q_{\alpha \beta} | \!\uparrow \rangle &=&
\left( \begin{array}{ccc} \
\frac{11}{21} & i & 0 \\
- i & \frac{11}{21} & 0 \\
0 & 0 & -\frac{22}{21} \end{array} \right), \\
\langle  \downarrow \!| \hat Q_{\alpha \beta} | \!\uparrow \rangle &=&
\left( \begin{array}{ccc}
\sqrt{\frac{14}{15}} & \sqrt{\frac{14}{15}} i & 0 \\
\sqrt{\frac{14}{15}} i &  -\sqrt{\frac{14}{15}} & 0 \\
0 & 0 & 0 \end{array} \right).
\end{eqnarray}
This gives rise to an XXZ model with the following parameters:
\begin{eqnarray}
\epsilon_0 &=& \frac{56}{15} , \\
\beta &=&  - \frac{351}{686}, \\
h_z &=& -\frac{1}{\epsilon_0 J_{ex}} \left(3 E_s - H_z - \frac{536}{147}
J_{ex} \right).
\end{eqnarray}

\subsubsection{$|2,2 \rangle \rightarrow |4,4\rangle$}
We define: $|\! \uparrow \rangle = |4,4\rangle$ and $|\! \downarrow \rangle  =|2,2 \rangle$. The relevant
non vanishing matrix-elements are:

\begin{eqnarray}
\langle  \uparrow \!| \hat Q_{\alpha \beta} | \!\uparrow \rangle &=&
\left( \begin{array}{ccc} \
\frac{2}{3} & 2 i & 0 \\
-2 i & \frac{2}{3} & 0 \\
0 & 0 & -\frac{4}{3} \end{array} \right) ,\\
\langle  \downarrow \!| \hat Q_{\alpha \beta} | \!\uparrow \rangle &=&
\left( \begin{array}{ccc}
\sqrt{\frac{12}{7}} & \sqrt{\frac{12}{7}} i & 0 \\
\sqrt{\frac{12}{7}} i &  -\sqrt{\frac{12}{7}} & 0 \\
0 & 0 & 0 \end{array} \right), \\
\langle  \downarrow \!| \hat Q_{\alpha \beta} | \!\downarrow \rangle &=&
\left( \begin{array}{ccc} \
\frac{11}{21} & i & 0 \\
- i & \frac{11}{21} & 0 \\
0 & 0 & -\frac{22}{21} \end{array} \right).
\end{eqnarray}
Using this we get again an effective Hamiltonian in the form of an XXZ-model. The parameters are:
\begin{eqnarray}
\epsilon_0 &=& \frac{48}{7} J \\
\beta &=& -\frac{284}{49} \frac{J_{ex}}{J} = - \frac{71}{84} \\
h_z &=& - \frac{1}{\epsilon_0 J_{ex}} \left(7 E_s - H_z - \frac{344}{49} J_{ex} \right)
\end{eqnarray}

\subsection{$N=3$}
For three particles per site we have the transition between $|\! \uparrow \rangle = |3,3\rangle$
and $|\! \downarrow \rangle = |1,1\rangle$. The non-vanishing matrix-elements are:
\begin{eqnarray}
\langle  \uparrow \!| \hat Q_{\alpha \beta} | \!\uparrow \rangle &=&
\left( \begin{array}{ccc} \
\frac{1}{2} & \frac{3}{2} i & 0 \\
-\frac{3}{2} i & \frac{1}{2} & 0 \\
0 & 0 & -1 \end{array} \right) \\,
\langle  \downarrow \!| \hat Q_{\alpha \beta} | \!\uparrow \rangle &=&
\left( \begin{array}{ccc}
\sqrt{\frac{3}{5}} & \sqrt{\frac{3}{5}} i & 0 \\
\sqrt{\frac{3}{5}} i &  -\sqrt{\frac{3}{5}} & 0 \\
0 & 0 & 0 \end{array} \right), \\
\langle  \downarrow \!| \hat Q_{\alpha \beta} | \!\downarrow \rangle &=&
\left( \begin{array}{ccc} \
\frac{3}{10} & \frac{i}{2} & 0 \\
-\frac{i}{2} & \frac{3}{10} & 0 \\
0 & 0 & -\frac{3}{5} \end{array} \right).
\end{eqnarray}
The parameters of the XXZ-model in this case are:
\begin{eqnarray}
\epsilon_0 &=& \frac{12}{5}  \\
\beta &=&  - \frac{8}{15} \\
h_z &=& - \frac{1}{\epsilon_0 J_{ex}} \left(5 E_s - H_z - \frac{124}{25}
J_{ex} \right)
\end{eqnarray}

\section{Effective Hamiltonian for Large $N$}
For large $N$, $N$ even, we can study all transitions from $|S, S\rangle
\rightarrow |S+2, S+2 \rangle$.
These states are given by:
\begin{equation}
|S,S\rangle = |Y_{SS} ({\bf n}) \rangle = (-1)^S \sqrt{\frac{2S+1}{4 \pi} \frac{(2S)!}{2^{2S} (S!)^2}}
 e^{i S \phi} \sin^S \theta
\end{equation}
The hopping term in the Hamiltonian is just equal to:
\begin{equation}
-2 J_{ex} \sum_{\langle  kl \rangle} ({\bf n}_k \cdot {\bf n}_l)^2
\end{equation}

Introducing $|\! \uparrow \rangle =|S+2,S+2\rangle$ and $|\! \downarrow \rangle =|S,S\rangle$,
we get the following non-vanishing matrix-elements:
\begin{eqnarray*}
\langle  \uparrow |_k \langle  \uparrow |_l ({\bf n}_k \cdot {\bf
n}_l)^2 | \uparrow \rangle_k | \uparrow \rangle_l
&=& \frac{19 + 12 S + 2 S^2}{(7+2 S)^2} \\
\langle  \uparrow |_k \langle   \downarrow |_l ({\bf n}_k \cdot {\bf n}_l)^2
| \uparrow \rangle_k | \downarrow \rangle_l
&=& \frac{7+8 S + 2 S^2}{(3+2S)(7+2S)} \\
\langle  \uparrow |_k \langle   \downarrow |_l ({\bf n}_k \cdot {\bf n}_l)^2
| \downarrow \rangle_k | \uparrow \rangle_l
&=& \frac{(2+S)(1+ S)}{(3+2S)(5+2S)} \\
\langle  \downarrow |_k \langle  \downarrow |_l ({\bf n}_k \cdot {\bf n}_l)^2 |
\downarrow \rangle_k | \downarrow \rangle_l
&=& \frac{3 + 4 S + 2 S^2}{(3+2 S)^2}
\end{eqnarray*}

The effective Hamiltonian turns out to be
\begin{eqnarray*}
{H}_{eff.} &=& -J_{ex} \sum_{ \langle  k l \rangle} \frac{(2+S)(1+S)}{(3+2
S)(5+2S)} (\sigma_k^x \sigma_l^x +
\sigma_k^y \sigma_l^y) \\
&& -J_{ex} \sum_{\langle  kl \rangle} \frac{12}{(3+2 S)^2(7+2 S)^2}
\sigma_k^z \sigma_l^z \\
&& -\left( (2 S+3) E_s - B_z - \frac{8 (3+ S)(1+ 2S)}{(3+ 2S)^2(7+ 2S)^2}
J_{ex} \right) \sum_k \sigma^z_k
\end{eqnarray*}
This is clearly an XXZ model with:
\begin{eqnarray}
\epsilon_0&=&\frac{(2+S)(1+S)}{(3+2 S)(5+2 S)}  \\
\beta &=& -1 + \frac{12 (5 +  2 S)}{((7 + 2 S)^2 (6 + 13 S + 9 S^2 + 2 S^3)} \\
h_z &=& - \frac{1}{\epsilon_0 J_{ex}} \left(( 2 S+3) E_s - B_z - \frac{8 (3+
S)(1+2S)}{(3+ 2S)^2(7+2S)^2} J_{ex} \right) \nonumber
\end{eqnarray}
It is clear that $\beta$ is negative for all $S$ and it get's more negative if we increase $S$.

For the lowest transition: $|0,0\rangle \rightarrow |2,2\rangle$ this gives the numbers:

\begin{eqnarray}
\epsilon_0 &=& \frac{2}{15} \\
\beta &=& -\frac{39}{49} \\
h_z&=&- \frac{1}{\epsilon_0 J_{ex}}\left( 3 E_s - H_z - \frac{8}{147} J_{ex}
\right)
\end{eqnarray}

\section{The Holstein-Primakov bosons Representation}

\subsection{Holstein-Primakov bosons in UP or DP phases}

The Hamiltonian of the XXZ-model is given as
\begin{eqnarray}
\frac{H_{XXZ}}{\epsilon_0 J_{ex}} = - \sum_{\langle  k l \rangle} \, \sigma_{k \alpha} \cdot
\sigma_{l \alpha} -
\beta \, \sum_{\langle  k l \rangle} \, \sigma_{k z}  \cdot \sigma_{lz} -
h_z \, \sum_k \sigma_{kz}.
\end{eqnarray}

In this subsection,
we are interested in Region I (See Fig.2) where $h_z+2d\beta >0$.
After Fourier transforming and in terms of H.P. bosons the Hamiltonian
can written as
\begin{equation}
H_{XXZ} =H^{(0)} + H^{(2)} +H^{(4)} + {\cal O}((\hat{c}^{(\dag)})^6).
\end{equation}

Here
\begin{equation}
H^{(0)} = -V^T \, ( d (\beta+1)  +  h_z);
\end{equation}
\begin{equation}
H^{(2)} = \sum_{\bf q} \, \left[ 4 (1+\beta)d + 2 h_z - 4 \sum_{\alpha} \cos ( q_\alpha \,
a) \right]
{c}^\dag_{\bf q} {c}_{\bf q};
\label{2nd}
\end{equation}
and the fourth order term is

\begin{eqnarray}
\frac{H^{(4)}}{J_{ex} \epsilon_0} &=& \, \frac{1}{V_T} \sum_{{\bf q}_1 {\bf q}_2 {\bf q}_3}
{c}^\dagger_{{\bf q}_1} {c}^\dagger_{{\bf q}_2} {c}_{{\bf q}_3} {c}_{{\bf q}_1+{\bf q}_2
-{\bf
q}_3} \nonumber \\
&&\sum_{\alpha} [ \exp (-i {\bf q}_{2\alpha} a) + \exp ( -i {\bf
q}_{3\alpha}a ) +
\nonumber \\
&& \exp ( -i (-{\bf q}_1-{\bf q}_2+{\bf q}_3)_\alpha a) + \exp(i {\bf
q}_{1\alpha} a) ] \nonumber \\
&&- 4 (1+\beta) \, \sum_{{\bf q}_1 {\bf q}_2 {\bf q}_3} \, {c}^\dagger_{{\bf q}_1} {c}^
\dagger_{{\bf
q}_2}
{c}_{{\bf q}_3} {c}_{{\bf q}_1+{\bf q}_2 -{\bf q}_3} \,  \nonumber\\
&&
\times \sum_{\alpha} \, \exp(i ({\bf q}_1 -{\bf q}_3)_\alpha a). \nonumber
\end{eqnarray}
Here $a$ is the lattice constant.

Following Eq.(\ref{2nd}), the energy of the quasi-particles is given by
\begin{eqnarray}
\epsilon_{\bf q} = 4 (1+\beta)  d  -
4\sum_{\alpha=x,y,z} \cos ({\bf q}_\alpha a) +2h_z,
\end{eqnarray}
where the energy gap in the spectrum is given as
\begin{equation}
\Delta(\beta,h_z) = 4\beta d + 2 h_z.
\end{equation}

The fourth order term describes interactions between magnons.
Indeed, in the small $|{\bf q}|$ limit
the hamiltonian can approximately be written as (up to a constant)
\begin{equation}
\frac{H_{XXZ}}{J_{ex} \epsilon_0} = \sum_{{\bf k}} \, \epsilon_{{\bf q}} 
{c}^\dagger_{{\bf q}}
\hat{c}_{{\bf q}}
-4 \beta d \, \sum_{{\bf q}_1 {\bf q}_2 {\bf q}_3} \, {c}^\dagger_{{\bf q}_1} {c}^\dagger_{{\bf
q}_2}
{c}_{{\bf
q}_3} {c}_{{\bf q}_1 + {\bf q}_2 -{\bf q}_3}.
\end{equation}
When $\beta>0$ interactions between the magnons are attractive and when
$\beta<0$ interactions are repulsive.

To derive these results, we have used the dilute gas expansion expansion
\begin{equation}
\sqrt{2S - c^\dagger c} =  \left(1 - \frac{c^\dagger c}{2} - \frac{1}{8} (c^\dagger c)^2
+\ldots \right).
\end{equation}

\subsection{Holstein-Primakov Bosons in Ferromagnetically Ordered Phase}

The most convenient way to study HP bosons in region III is to introduce the following rotation:
\begin{equation}
\left( \begin{array}{c} x \\ y \\ z \end{array} \right) =
\left( \begin{array}{c} cos \Theta x' + \sin \Theta z' \\ y' \\ cos \Theta x ' - \sin \Theta z ' \end{array} \right)
\end{equation}
In the semiclassical approximation,
by minimizing the energy with respect to $\Theta$, one obtains the ground state solution
with $\cos \Theta = - \frac{h_z}{2 d \beta }$.

Consider an expansion over this solution.
We get the following lowest order terms:
\begin{eqnarray}
\frac{H_{XXZ}^{(2)}}{\epsilon_0 J_{ex}}
 &=& \sum_{\bf q} \left(4d + 4 d \beta \cos^2 \Theta + 2 {h_z} \cos \Theta\right) c_{\bf q}^\dagger c_{\bf q} \\&&
 - \sum_{\bf q} \left(4 + 2 \beta \sin^2 \Theta \right) \sum_\alpha \cos {\bf q}_\alpha a \;  c_{\bf q}^\dagger c_{\bf q} \nonumber \\ &&
 -  \sum_{\bf q}  {\beta} \sin^2 \Theta  \sum_\alpha \cos {\bf q}_\alpha a \left( c_{\bf q} c_{\bf q} + c_{\bf q}^\dagger c_{\bf q}^\dagger \right)
\nonumber
\end{eqnarray}
When $\Theta=0$, one recovers the results in section V.B.

Taking into account $\cos \Theta = - \frac{h_z}{2 d \beta }$ in the ferromagnetic phase,
in the long wave length limit one further simplies the result to
\begin{eqnarray}
\frac{{H}_{XXZ}^{(2)}}{\epsilon_0 J_{ex}} &=&  \sum_{\bf q}  4(d-\sum_\alpha  \cos {\bf q}_\alpha a)
c_{\bf q}^\dagger c_{\bf q} \\&&
- \sum_{\bf q} \left( 2 d \beta \sin^2 \Theta \right) \;  c_{\bf q}^\dagger c_{\bf q} \nonumber \\ &&
- \sum_{\bf q} d {\beta} \sin^2 \Theta  \left( c_{\bf q} c_{\bf q} + c_{\bf q}^\dagger c_{\bf q}^\dagger \right) \nonumber.
\end{eqnarray}

This yields the following dispersion
\begin{eqnarray}
\omega_{\bf q}
&=& 2\sqrt{2} a \sqrt{-{ d\beta}} \sqrt{1 - \frac{h_z^2}{4 d^2 \beta^2}} {|\bf q|}.
\label{sm}
\end{eqnarray}
eq.(\ref{sm})
agrees with the results derived in the dilute gas approximation
in section V.C; close to critical lines, one notices that $\sin\Theta=2\sqrt{n_0}$ and $d\beta \sin^2\Theta= 4 d \beta n_0$.

\end{multicols}


\begin{thebibliography}{99}



\bibitem{Haldane83}F. D. M. Haldane, Phys. Rev. Lett. {\bf 50},
1153 (1983).

\bibitem{Affleck87}I. Affleck, T. Kennedy, E. H. Lieb and H. Tasaki,
Phys. Rev. Lett. {\bf 59}, 799 (1987).


\bibitem{Arovas88}D. P. Arovas, A. Auerbach and  F. D. Haldane, Phys. Rev.
Lett. {\bf 60}, 531 (1988).



\bibitem{Affleck91}I. Affleck, Phys. Rev. {\bf B} 43, 3215 (1991).

\bibitem{Tsvelik90}In one-dimensions, a Majorana fermion representation was employed to
study magnetization in A. Tsvelik, Phys. Rev. {\bf B 42}, 
10499 (1990).


\bibitem{Nomura91}K. Nomura and T. Sakai, Phys. Rev. {\bf B 44}, 5092 (1991);
T. Sakai and M. Takahashi, Phys. Rev. {\bf B43}, 13383 (1991).

\bibitem{Sorensen93}E. Sorensen and I. Affleck, Phys.Rev.Lett.{\bf 71},1633 (1993).


\bibitem{Nikuni00}T. Nikuni, M. Oshikawa, A. Oosawa and H. Tanaka,
Phys. Rev. Lett. {\bf 84}, 5868 (2000).

\bibitem{Mishguich04}G. Mishguich and M. Oshikawa, cond-mat/0405422.













\bibitem{Demler02}E. Demler and F. Zhou, Phys. Rev. Lett. {\bf 88}, 
163001-1 (2002); E. Demler, F. Zhou and D. F. M. Haldane, 
ITP-UU-01/09 (2001).


\bibitem{Zhou03a}F. Zhou and M. Snoek,
Annals of Physics {\bf 308}, 692 (2003).




\bibitem{Imambekov03}
A. Imambekov, M. Lukin
and E. Demler, Phys. Rev. {\bf A68}, 063602 (2003).

\bibitem{Snoek04}M. Snoek and F. Zhou, Phys. Rev. {\bf B 69}, 094410 (2004).




\bibitem{Zhou03b}Fei Zhou,
Euro. Phys. Lett. {\bf 63}(4), 505 (2003)
[See also cond-mat/0207041].



\bibitem{Wiemer03}J. Wiemer and F. Zhou, cond-mat/0309312.


\bibitem{Ho00}
T. L. Ho, Phys. Rev. Lett. {\bf 81}, 
742 (1998); 
C. K. Law, P. Han and N. Bigelow, Phys. Rev. Lett. {\bf 81}, 742 (1998);
T. Ohmi and K. Machida, J. Phys. Soc. Jpn. {\bf 67}, 1822 (1998).
Discussions about the biquadratic Zeeman effects can be found in
T. L. Ho and S. Yip, Phys. Rev. Lett. {\bf 84}, 4031 (2000). 






\bibitem{Stamper-Kurn98}
For experiments on spinor condensates, see
D. M. Stamper-Kurn, M. R. Andrews, A. P. Chikkatur, S.
Inouye, H.-J. Miesner, J. Stenger and W. Ketterle, 
Phys. Rev. Lett. {\bf 80}, 2027 (1998); J. Stenger, S. Inouye, D. M. Stamper-Kurn, H.-J.
Miesner,
A. P. Chikkatur and W. Ketterle, Nature {\bf 396}, 345 (1998);






\bibitem{Imambekov04}A. Imambekov, M. Lukin and E. Demler, cond-mat / 0401526.

\bibitem{Zhou01}F. Zhou, Phys. Rev. Lett. {\bf 87}, 080401-1 (2001);
F. Zhou,
Int. Jour. Mod. Phys. {\bf B 17} No. 14,
2643-2698 (2003) [also cond-mat/0108473].



\bibitem{Mueller03}This traceless projected order parameter differs from the characterization discussed in
E. Mueller, cond-mat/0309511.



\bibitem{hd}The dilute gas approximation applies only in high dimensional lattices.
In one-dimensions, fermionization is needed and this subject will be examined in
a subsequent work.


\bibitem{Nozieres}P. Nozieres and D. Pines,
{\em The theory of Quantum Liquids, Vol II: Superfluid Bose Liquids},
Addison-Wesley Co., Inc (1990).





















\end{thebibliography}
\end{document}